\documentclass[a4paper,11pt]{article}

\usepackage{jheppub} 

\usepackage{empheq,mathtools}
\usepackage{amsmath,amsfonts,amssymb}
\usepackage{mathrsfs}
\usepackage{cancel,slashed}
\usepackage{tensor}
\usepackage{xcolor}
\usepackage{graphicx}
\usepackage{subfigure}
\usepackage{float}

\makeatletter
\def\@fpheader{\relax}
\makeatother

\usepackage{ulem}
\usepackage[czech,english]{babel}
\usepackage{yfonts}
\usepackage{url}
\usepackage{mathabx}

\newcommand{\be}{\begin{equation}}
\newcommand{\ee}{\end{equation}}

\preprint{PCFT-24-32,  LCTP-24-18}

\title{Quantum Corrections to Holographic Strange Metal at Low Temperature}

\author[a]{Xiao-Long Liu,}
\author[a,b]{Jun Nian,}
\author[c,d]{Leopoldo A. Pando Zayas }

\affiliation[a]{International Centre for Theoretical Physics Asia-Pacific,\\ University of Chinese Academy of Sciences, 100190 Beijing, China}
\affiliation[b]{Peng Huanwu Center for Fundamental Theory, Hefei, Anhui 230026, China}
\affiliation[c]{Leinweber Center for Theoretical Physics, University of Michigan, Ann Arbor, MI 48109, USA}
\affiliation[d]{The Abdus Salam International Centre for Theoretical Physics, 34014 Trieste, Italy}

\emailAdd{liuxiaolong22@mails.ucas.ac.cn}
\emailAdd{nianjun@ucas.ac.cn}
\emailAdd{lpandoz@umich.edu}

\abstract{The holographic approach to the strange metal phase relies on near-extremal asymptotically AdS$_4$ electrically charged black branes with important input from their AdS$_2$ near-horizon throat geometry. Motivated by the current understanding of the role of quantum fluctuations in the throat of near-extremal black holes, we revisit some transport properties. We model quantum gravitational and gauge fluctuations in the throat region by adopting results in Jackiw-Teitelboim gravity, effectively leading to quantum corrections for the dual CFT$_1$ Green's function in the near-horizon infrared region. We use the quantum-corrected Green's function to compute the conductivity for (2+1)-dimensional holographic strange metals and obtain corrections for the DC resistivity and the optical conductivity. We also compare the quantum-corrected holographic approach with results from the complex Sachdev-Ye-Kitaev model and point out qualitative differences. Although experimental detection for the quantum-corrected holographic approach to the DC resistivity requires higher precision than current experimental accuracy, future experiments with improved technologies could detect these quantum corrections. Interestingly, including quantum corrections to the optical conductivity does provide a plausible explanation for the experimental anomalous power-law behavior detected in various strange metals.}

\begin{document} 
\maketitle

\section{Introduction}

\begin{figure}[htb!]
\begin{center}
\includegraphics[width=10cm, angle=0]{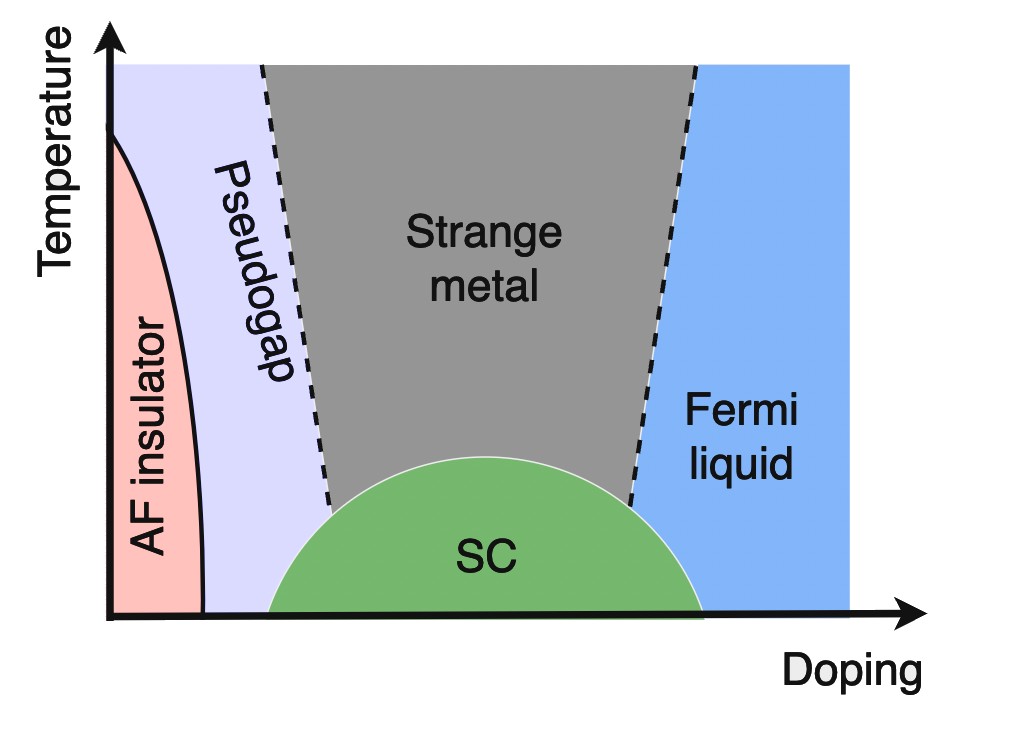}
\caption{The phase diagram of cuprates (see e.g. \cite{Banerjee2018HighTS, articleVarma}).}\label{fig:PhaseDiagram}
\end{center}
\end{figure}

Strange metals were discovered in conjunction with the study of high-temperature superconductivity in a class of copper oxides (see the phase diagram Fig.~\ref{fig:PhaseDiagram}). Strange metals derived their name from the anomalous behavior of some of their transport properties, including DC conductivity, optical conductivity, Hall angle, etc.; understanding these properties is a long-standing puzzle in the theory of strongly correlated electron systems \cite{PhysRevLett.69.2975, Bruin:2013qlp, PhysRevB.96.155449, Brown:2019vef} (see also  reviews \cite{Hartnoll:2021ydi,Phillips:2022nxs}). The paradigmatic anomalous behavior for strange metal is the linear resistivity for a large range of temperatures  (${\rm mK} - 1000\;{\rm K}$).  Understanding this linear resistivity is arguably crucial for deciphering the mechanism of high-temperature superconductivity. However, after over three decades, there is still no entirely satisfactory explanation.

Besides the linear resistivity, there are various other anomalous properties for strange metals, including an anomalous power-law dependence of optical conductivity, $\sigma(\omega)\sim \omega^{-\alpha}$ for $T \to 0$, which is different from the standard Drude behavior and has been observed in cuprates \cite{Baraduc1996InfraredCI, PhysRevB.49.9846, Hwang2006DopingDO, vandeMarel:2003wn}. Sometimes, this law can also be formulated as $\sigma(\omega,T) \sim \omega^{-\alpha}\, F(\omega,T)$ \cite{PhysRevB.66.041104, PhysRevB.87.035102, Michon:2023qpl, Marel2006ScalingPO}, where $F(\omega,T)$ contains some correction terms of higher orders in $\omega$. The most typical value of $\alpha$ for strange metals is 2/3 \cite{PhysRevB.58.11631, Marel2006ScalingPO, Phillips:2022nxs}, but some other values have also been observed in experiments (e.g., 0.5 \cite{PhysRevB.66.041104}, 0.6 \cite{vandeMarel:2003wn}, 0.65 \cite{vandeMarel:2003wn}, 0.7 \cite{Baraduc1996InfraredCI}, 0.77 \cite{PhysRevB.49.9846}, 0.8 \cite{Michon:2023qpl}, etc.).

To explain the properties of strange metals, various models in condensed matter physics have been proposed, such as the Sachdev-Ye-Kitaev (SYK) model \cite{Sachdev:1992fk, Kitaev, Guo:2020aog, RevModPhys.94.035004}, the Yukawa-SYK model \cite{PhysRevB.106.115151, Patel:2022gdh, Li:2024kxr, Sachdev:2024iux}, and the marginal Fermi liquid (MFL) \cite{Varma:1989zz}, which have led to partially satisfactory answers.

It is worth recalling that strange metallicity, particularly in cuprates,  is often conceptually approached from the notion of Planckian dissipation where the physics is controlled by a universal relaxation time $\tau = \frac{\hbar}{k_B\, T}$. The absence of system parameters insinuates a form of scale invariance, thus inviting holographic approaches.

In this manuscript, we will be concerned with the strange metal phase within the AdS/CFT correspondence. Over the last three decades, developments in string theory brought about the anti–de-Sitter/conformal field theory (AdS/CFT) correspondence \cite{Maldacena:1997re, Gubser:1998bc, Witten:1998zw}. In its simpler form, the AdS/CFT correspondence relates a weakly coupled gravity theory in ($d$+1)-dimensional AdS spacetime with a strongly coupled $d$-dimensional quantum field theory on the boundary. Thus, this correspondence transforms questions about complex many-body quantum phenomena at strong coupling into single- or few-body classical problems in a curved geometry. Indeed, the tools of the AdS/CFT correspondence were brought to bear on the question of anomalous properties of strange metals in a beautiful series of works \cite{Faulkner:2013bna, Faulkner:2010zz, Faulkner:2009wj, Faulkner:2010da, Faulkner:2011tm, Liu:2009dm} which reported a class of non-Fermi liquids. These works focused on a (2+1)-dimensional boundary theory, working with a Reissner-Nordstr\"om (RN) black brane in an asymptotically AdS$_4$ spacetime, and took the boundary theory to be conformally invariant. This model recovered the linear resistivity behavior and some other properties of strange metals. It is worth pointing out that the black brane solution used in the above works corresponds to a low-temperature background, which gravitationally is denoted as a near-extremal background, and it is known that such solutions develop an infinite AdS$_2$ throat in their near-horizon region. This AdS$_2$ region featured prominently in the computations of \cite{Faulkner:2013bna, Faulkner:2010zz, Faulkner:2009wj, Faulkner:2010da, Faulkner:2011tm, Liu:2009dm}.

There has recently been impressive progress in the understanding of the quantum nature of the AdS$_2$ region 
\cite{Almheiri:2014cka, Jensen:2016pah, Maldacena:2016upp}. The progress was originally achieved in the simple framework of Jackiw-Teitelboim gravity (JT) \cite{Jackiw:1984je, Teitelboim:1983ux}. The understanding of the nature of the quantum fluctuations was later applied to higher-dimensional near-extremal black holes and led to important modifications to the low-temperature thermodynamics of black holes \cite{Iliesiu:2020qvm, Heydeman:2020hhw, Boruch:2022tno, Iliesiu:2022onk}. A similar approach led to a resolution of an old puzzle related to the entanglement entropy of holographic quantum fields in Rindler space \cite{Emparan:2023ypa}. More recently, the physics of the throat region is essential in leading to logarithmic-in-temperature corrections to the thermodynamics of rotating black holes in asymptotically flat spacetimes \cite{Kapec:2023ruw, Rakic:2023vhv}. These results were universal in \cite{Maulik:2024dwq}, extending the one-loop quantum corrections to the thermodynamics for asymptotically AdS spacetimes and various dimensions. The corrections to the thermodynamics are obtained by regulating a set of zero modes following recent suggestions of turning on a small but finite temperature in the geometry \cite{Iliesiu:2022onk, Banerjee:2023quv, Banerjee:2023gll}. Other related explorations have been reported recently \cite{Bai:2023hpd, Kapec:2024zdj, Kolanowski:2024zrq}.

Collectively, many of these works have demonstrated that temperature acts as a sort of coupling, with high temperature corresponding to the classical regime and low temperature to the quantum, strongly coupled regime. The main purpose of this manuscript is to bring the current understanding of quantum fluctuations in the throat region to bear on previous results obtained using AdS$_4$/CFT$_3$, where a throat region plays a prominent role, such as in \cite{Faulkner:2013bna, Faulkner:2010zz, Faulkner:2009wj, Faulkner:2010da, Faulkner:2011tm, Liu:2009dm}. More concretely, we compute the effects of quantum corrections in the throat on physical observables of holographic strange metal at low temperatures, such as DC and optical conductivities. We will also compare the holographic approach with the direct treatment of strange metal using the complex Sachdev-Ye-Kitaev model.  We will highlight the possibility of experimentally detecting the influence of these quantum corrections.

Let us describe the main approximations we take in this paper. We start with the main AdS/CFT formula
\begin{equation}
    Z_{\rm CFT}[J]=Z_{\rm bulk}[\phi_0],
\end{equation}
where $\phi_0$ is the boundary value of the bulk field dual to the operator that couples to the source $J$. The holographic computation of the (2+1)-dimensional Green's function, $G_R^{3d}$, is connected to the (0+1)-dimensional retarded Green's function of the throat, $\mathcal{G}_R^{1d}$, via \eqref{eq:ads4green}. We will approximate the retarded Green's function of the throat by using the coupling in the throat region to be $\varphi_0$. Namely,
\begin{eqnarray}
   \frac{\delta^2 Z_{\rm JT}^{2d}[\varphi_{0}]}{\delta \varphi_0^2 } &=& \frac{\delta^2}{\delta\varphi_0^2} \int [Dg_{\mu\nu}][D\phi]\, \exp\Big[-S(g_{\mu\nu},\phi;\varphi_0)\Big]\nonumber \\
    &\approx & \frac{\delta^2}{\delta\varphi_0^2} \int [D f]\,  \exp\Big[-S_{\rm 2d}(f(t)) -S_{\rm matter}[\varphi_0]\Big] \nonumber\\
   &=& \int [Df]\, e^{-S_{2d}[f(t)]}\, G_R(t) = \langle G_R (t)\rangle\, ,
\end{eqnarray}
where $S_{\rm 2d}$ denotes the two-dimensional action, which contains the Schwarzian and the action for the gauge fluctuations. It is this quantum-corrected throat-determined Green's function that will enter in determining the (2+1)-dimensional retarded Green's function.  Due to a noted property at very low frequencies, the 1d Green's function feeds into the full computation of the 3d Green's function  \cite{Faulkner:2009wj}. This improves our approximation because it is quite reasonable to expect that the JT physics completely dominates in the throat.

This paper is organized as follows. In Sec.~\ref{sec:Review}, we review the holographic model of strange metal in the literature without quantum corrections from the throat region. Next, we discuss how to introduce quantum corrections in the AdS$_4$/CFT$_3$ framework in Sec.~\ref{sec:AdS4CFT3}, where explicit results of quantum-corrected DC and optical conductivities will be provided. In Sec.~\ref{sec:CplxSYK}, we review the quantum corrections as they appear in the framework of the (0+1)-dimensional complex SYK model and note that they affect the physical observables (DC and optical conductivities) in a qualitatively different way. In Sec.~\ref{sec:Discussion}, we provide a summary of the results and discuss a sweet spot scenario where the corrections introduced in this paper could be detected experimentally. Some computational details are presented in the Appendices.

\section{ Review of the AdS$_4$/CFT$_3$ holographic model for strange metal}\label{sec:Review}

\subsection{Gravity setup}

Let us describe the minimal ingredients required on the gravity side to help us describe the field theory physics we are interested in. To describe conductivity, we need, beyond the metric tensor, a bulk $U(1)$ gauge field, which will be dual to a global $U(1)$ symmetry in the dual field theory. The corresponding Einstein-Maxwell action with a negative cosmological constant is 
\begin{equation}\label{eq:action1}
S=\frac{1}{2\kappa^2}\int d^4x\sqrt{-g}\left(\mathcal{R}+\frac{6}{R^2}-\frac{R^2}{g^2_F}F_{MN}F^{MN}\right)\, ,
\end{equation}
where $R$ is the radius of AdS$_4$, and $g_F$ is an effective dimensionless gauge coupling. Let us recall a few facts from the AdS/CFT dictionary. The dual CFT$_3$ is defined on the boundary of AdS$_4$. We consider a finite density state for CFT$_3$ with a chemical potential $\mu$ for an $U(1)$ global symmetry. The associated conserved current $J_{\mu}$ of the boundary global $U(1)$ is coupled to the bulk gauge field $A_M$. The central power of the AdS/CFT correspondence resides in providing insights into patterns of symmetry breaking and universality in strongly coupled field theories via gravity manipulations. Ultimately, we use this framework to approximate a boundary CFT$_3$ capable of describing generic aspects of the (2+1)-dimensional strange metal phase.

\begin{figure}[htb!]
\begin{center}
\includegraphics[width=11cm, angle=0]{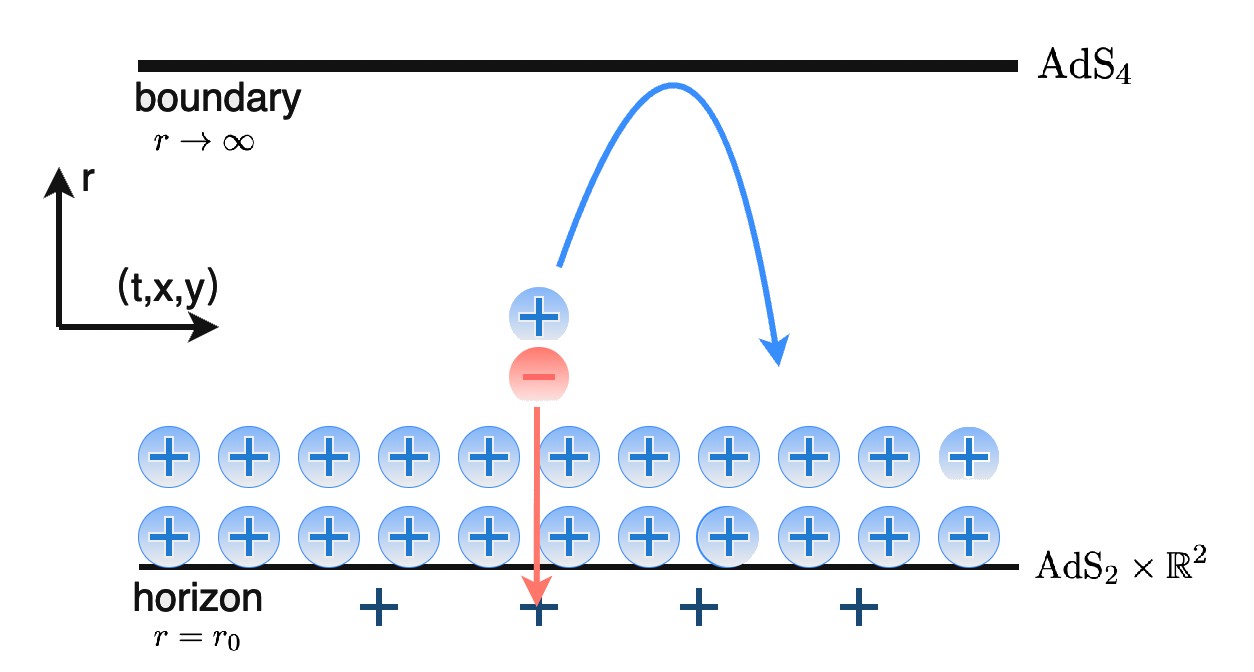}
\caption{The mechanism of holographic strange metal as  depicted in \cite{Faulkner:2010da,Faulkner:2010zz}.}\label{fig:background}
\end{center}
\end{figure}

The equations of motion from the action \eqref{eq:action1} admits the charged black brane geometry as a solution \cite{Chamblin:1999tk,Romans:1991nq}:
\begin{align}
ds^2 & = \frac{r^2}{R^2}\left[-f(r)dt^2 + dx^2_i\right] + \frac{R^2}{r^2}\frac{dr^2}{f(r)}\, ,\label{eq:metricads4}\\
  A_t & = \mu \left(1-\frac{r_0}{r}\right)\, ,\label{eq:gaugeAdS4}
\end{align}
with
\begin{equation}
f(r)=1+\frac{Q^2}{r^4}-\frac{M}{r^3},
\end{equation}
where $r_0$ is the horizon radius. The Hawking temperature, chemical potential, density of charge, energy, and entropy density are given by
\begin{align}
 T & = \frac{3r_0}{4\pi R^2} \left(1-\frac{Q^2}{3r_0^4}\right),\quad \mu = \frac{g_F\, Q}{R^2r_0},\quad \rho = \frac{2Q}{\kappa^2R^2g_F},\nonumber\\
\epsilon & = \frac{1}{\kappa^2}\frac{M}{R^4},\quad s = \frac{2\pi}{\kappa^2} \left(\frac{r_0}{R}\right)^2\, .
\end{align}
To explore the response of a generic fermionic operator, $\mathcal{O}$,  in the dual field theory,  one considers a bulk spinor field $\psi$ with charge $q$ and mass $m$. The conformal dimension, $\Delta$, of $\mathcal{O}$ is related to $m$ by $\Delta=\frac{3}{2}+mR$. Quanta for the bulk field $\psi$ can be paired-produced near the horizon when $q$ is large compared to $mR$. Then, the negatively charged particle will fall into the horizon, and the positive one will move to the boundary of AdS$_4$. However, due to the curvature of AdS spacetime, it will fall back to the black hole, either falling into the horizon or being scattered by the black hole toward the boundary again. This process will finally reach an equilibrium, left with a finite density positively charged gas hovering outside the horizon, as shown in Fig.~\ref{fig:background}, borrowed from the original presentation in \cite{Faulkner:2010da}. The finite density of fermions corresponds to a finite density of excitations of the operator $\mathcal{O}$ in the boundary CFT.

It is appropriate to treat the bulk spinor field in the probe approximation. Indeed, note that the charge density carried by the black hole is determined by the classical geometry and background fields, which leads to a boundary theory density of the order $O(G^{-1}_N)$. In contrast, the density for $\psi$ quanta is of order $O(1)$. Thus, for classical geometry and background fields in the bulk, what we consider is just a small part of a large system.

\subsection{Green's functions in IR and UV CFTs}

In this subsection, we briefly review the traditional approach to studying the transport properties in the system of interest. 

Because Green's function in the UV CFT can be built up from that of the IR CFT, we will first discuss the problem in the near-horizon region. To probe the near-horizon region of the black brane, we consider the following scaling limit:
\begin{align}\label{eq:ads2limit}
r-r_*=\lambda\frac{R^2_2}{\zeta},\qquad r_0-r_*=\lambda\frac{R^2_2}{\zeta_0},\qquad t=\lambda^{-1}\tau,
\end{align}
factotaking $\lambda\to 0$ with $\zeta$, $\zeta_0$, and $\tau$ finite, where $r_* \equiv (Q / \sqrt{3})^{1/2}$, and $R_2$ is the AdS$_2$ radius. The scaling \eqref{eq:ads2limit} defines a new variable $\zeta$ and a new length scale $\zeta_0$. In this near-horizon limit, the metric \eqref{eq:metricads4} becomes AdS$_2\times\mathbb{R}^{2}$:
\begin{equation}\label{eq:metricads2}
ds^2=\frac{R^2_2}{\zeta^2} \left[- \left(1-\frac{\zeta^2}{\zeta^2_0}\right) d\tau^2 +\frac{d\zeta^2}{1-\frac{\zeta^2}{\zeta^2_0}} \right] + \frac{r^2_*}{R^2}d\vec{x}\,^2\, ,
\end{equation}
and the gauge field \eqref{eq:gaugeAdS4} becomes
\be
  A_{\tau}=\frac{g_F}{2\sqrt{3}} \left(\frac{1}{\zeta}-\frac{1}{\zeta_0}\right)\, .
\ee
Consequently, the Hawking temperature is $T=\frac{1}{2\pi\zeta_0}$, and the AdS$_2$ radius is $R_2=R/\sqrt{6}$. In the near-horizon AdS$_2$, one can also consider the action for a spinor field $\Psi$:
\begin{equation}
S=\int d^2x\sqrt{-g}\, i\, \left(\bar{\Psi}\Gamma^{\alpha}D_{\alpha}\Psi-m\bar{\Psi}\Psi+i\tilde{m}\bar{\Psi}\,\Gamma\,\Psi\,\right)\, ,
\end{equation}
where we introduce a time-reversal violating mass term proportional to $\tilde{m}$. By choosing the Gamma matrices $\Gamma^{\underline{\zeta}}=\sigma^3,\;\Gamma^{\underline{\tau}}=i\sigma^1,\; \Gamma = -\sigma^2$, we have the Dirac equation in the AdS$_2$ background \eqref{eq:metricads2}. Performing the Fourier transform $\Psi(\tau,\zeta) = \int d\omega\, e^{-i\omega\tau}\Psi(\omega,\zeta)$, we can express the Dirac equation in the frequency space:
\begin{equation}
\left(\partial_{\zeta} - i\sigma^3 \frac{\omega + q A_{\tau}}{\bar{f}}\right) \tilde{\Phi} = \frac{R_2}{\zeta \sqrt{\bar{f}}} \left(m \sigma^2 + \tilde{m} \sigma^1\right) \tilde{\Phi}\, ,
\end{equation}
where $\tilde{f} \equiv 1-\zeta^2/\zeta^2_0$ and $\tilde{\Phi} \equiv \frac{1}{\sqrt{2}}(1+i\sigma^1)\, \left(-gg^{\zeta\zeta}\right)^{-1/4}\Psi$. By solving this equation and following the standard procedure of the AdS/CFT correspondence \cite{Gubser:1998bc, Witten:1998zw}, we obtain the fermionic operator retarded Green's function in the near-horizon boundary CFT$_1$ (i.e., the IR CFT$_1$):
\begin{equation}\label{eq:ads2green}
\mathcal{G}_R(\omega,T)=(2\pi T)^{2\ell-1}\, g \left(\frac{\omega}{T},\frac{k}{\mu}\right)\, ,
\end{equation}
where 
\begin{align}
\ell & \equiv \frac{1}{\sqrt{6}}\sqrt{m^2R^2+\frac{3k^2}{\mu^2}-\frac{q^2}{2}}+\frac{1}{2}\, ,\\
g \left(\frac{\omega}{T},\frac{k}{\mu}\right) & \equiv \frac{\Gamma(-2\ell+1)}{\Gamma(2\ell-1)}\frac{\Gamma(\ell-\frac{i\omega}{2\pi T}+iqe_d)}{\Gamma(1-\ell-\frac{i\omega}{2\pi T}+iqe_d)}\frac{\Gamma(\ell+\frac{1}{2}-iqe_d)}{\Gamma(\frac{3}{2}-\ell-iqe_d)}\nonumber\\
 & \quad\times \frac{(m-i\frac{kRr_0}{r^2_*})R_2-iqe_d-\ell+\frac{1}{2}}{(m-i\frac{kRr_0}{r^2_*})R_2-iqe_d+\ell-\frac{1}{2}}\, ,
\end{align}
with $e_d \equiv g_F / 2\sqrt{3}$. We have taken $\tilde{m}=k R r_0 / r^2_*$, $g_F=1$, and $k = |\,\vec{k}\,|$, where $\vec{k}$ is the fermion momentum.

More generally, we can solve the Dirac equation in the full AdS$_4$ black hole background \eqref{eq:metricads4} and obtain the retarded Green's function $G_R^{3d}(\omega, T)$ for fermionic operator $\mathcal{O}$ in the boundary CFT$_3$ (i.e., the UV CFT$_3$). The behavior of $G_R^{3d}(\omega,T)$ near Fermi surface is given in terms of the near-horizon ${\cal G}_R$ as follows \cite{Faulkner:2009wj}:
\begin{equation}\label{eq:ads4green}
G_R^{3d}(\omega,k)=\frac{h_1}{k-k_F(\omega,T)-\Sigma(\omega,k)}\, ,
\end{equation}
where $\Sigma(\omega,T) = h_2\,\mathcal{G}(\omega,T)$, and $k_F(\omega,T)$ is approximately the Fermi momentum $k_F$ for low $\omega$ and $T$, while $h_{1, 2}$ are positive constants which can be fixed numerically. We see that the Green's functions in the UV CFT$_3$, $G_R^{3d}$, and in the IR CFT$_1$, $\mathcal{G}_R$, are related \cite{Faulkner:2009wj}. For the framework corresponding to rotating AdS black hole background, similar relations between the boundary CFT$_d$ and the near-horizon CFT$_2$ have also been studied \cite{Nian:2020bzf}. The UV CFT$_3$ Green's function $G_R^{3d}$ is directly related to the spectral density, which can subsequently be related to a bulk-to-bulk spinor propagator \cite{Faulkner:2013bna}.

\begin{figure}[htb!]
\begin{center}
\includegraphics[width=11cm, angle=0]{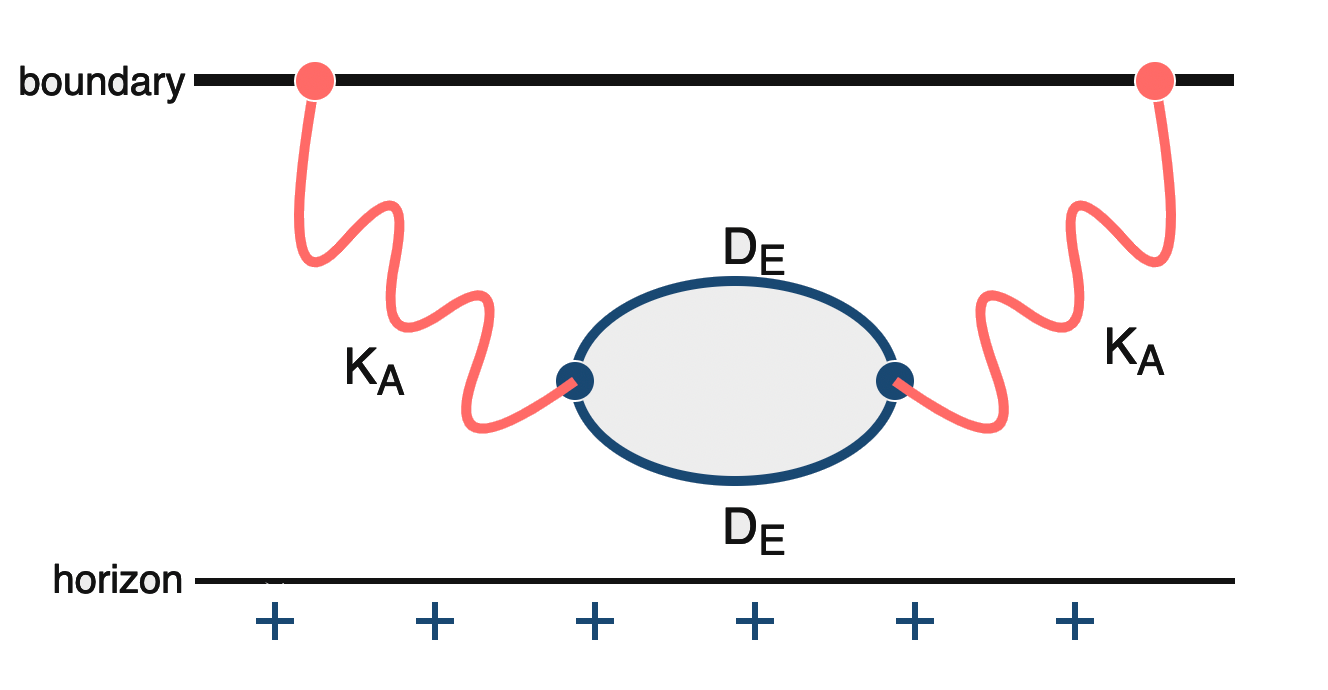}
\caption{The 1-loop Witten diagram contributing to the conductivity of strange metal at the order $O(N^0)$ as described in \cite{Faulkner:2010zz, Faulkner:2010da, Faulkner:2013bna}.}\label{fig:loop1}
\end{center}
\end{figure}

In a strange metal, the hole system contains a charged background of order $O(N^2)$. Thus, the conductivity for the hole system is divergent since there is no mechanism for current dissipation. We focus on the low-temperature contribution of order $O(N^0)$ from the Fermi surface. As explained in \cite{Faulkner:2010zz}, this contribution to the conductivity corresponds to a 1-loop Witten diagram shown in Fig.~\ref{fig:loop1}.

Applying the AdS/CFT correspondence, we can evaluate this 1-loop Witten diagram to obtain the Green's function in the boundary CFT$_3$:
\begin{align}
\langle J_y(\Omega)J_y(-\Omega)\rangle & \sim\int d\vec{k}\, dr_1\, dr_2\, D_E(r_1,r_2;i (\Omega + \omega),\vec{k})K_A(r_1;i\Omega)\nonumber\\
 & \qquad\times D_E(r_2,r_1;i\omega,\vec{k})K_A(r_2;-i\Omega)\, ,
\end{align}
where $D_E$ and $K_A$ are the bulk-to-bulk spinor and the boundary-to-bulk gauge propagators, respectively, and $\Omega$ is the momentum carried by the boundary current operator. The IR CFT$_1$ Green's function is encoded in the bulk-to-bulk spinor propagator $D_E$.  The quantum corrections to $D_E$ motivated by the new understanding of the pattern of symmetry breaking in the AdS$_2$ region and their implications for transport properties are the main new contributions of this manuscript. We start this discussion in the next section.

\subsection{DC and optical conductivities}

To compute the conductivity, we should follow the Kubo formula
\begin{equation}\label{eq:conductivity}
\sigma \equiv \sigma(\Omega) = \frac{1}{i\Omega}\langle J_y(\Omega)J_y(-\Omega)\rangle = \frac{1}{i\Omega} G^{yy}_R\, ,
\end{equation}
where $J_y$ is the current density in the $y$-direction on the boundary. Since the boundary has rotational symmetry, the choice of the $y$-direction is arbitrary. The conductivity can be written in terms of the boundary theory quantities \cite{Faulkner:2010zz, Faulkner:2013bna}:
\begin{align}
\sigma & = \frac{D}{i\omega}\int d\vec{k}\,\int\frac{d\omega_1}{2\pi}\frac{d\omega_2}{2\pi}\frac{f(\omega_1)-f(\omega_2)}{\omega_1-\omega-\omega_2-i\epsilon}\, A(\omega_1,\vec{k})\, \Lambda(\omega_1,\omega_2,\omega,\vec{k})\, \Lambda(\omega_2,\omega_1,\omega,\vec{k})\, A(\omega,\vec{k})\, ,
\end{align}
where $A(\omega,T) = {\rm Im}\, G(\omega,T) / \pi$ is the spectral density, $f(\omega)$ is the Fermi-Dirac distribution, $\Lambda$ is an effective vertex, and $D$ is a temperature-independent constant. In principle, $\Lambda$ is a complex function of $(\omega, \vec{k})$. However, when we focus on the low-temperature limit and near the Fermi surface, $\Lambda$ becomes a smooth real function of $|\,\vec{k}\,|$, which is independent of $(\omega, T)$. Thus, the conductivity is controlled only by $A(\omega, \vec{k})$, which depends on $\Sigma(\omega, k)$ and subsequently on the IR CFT$_1$ Green's function $\mathcal{G}(\omega, T)$. Finally, the DC conductivity at the leading order in small $T$ is given by
\begin{equation}\label{eq:dc1}
  \sigma_{DC}=\alpha T^{1-2\ell}\, ,
\end{equation}
where $\alpha$ is a numerical constant. For $\ell=1$, this corresponds to the marginal Fermi liquid (MFL), and one can recover the linear resistivity behavior for strange metal.

In contrast to the DC conductivity, the optical conductivity can be studied in the limit $\omega\gg T$ of the MFL and has the following expression at low frequency \cite{Faulkner:2013bna, Faulkner:2010zz}:
\begin{equation}
\sigma\sim\frac{ic}{\omega} \left[\frac{1}{{\rm log}\omega}+\frac{1}{({\rm log}\omega)^2}\frac{1+i\pi}{2}\right] + \cdots\, ,
\end{equation}
where $c$ is a negative constant. Experimental data have also confirmed the behavior of the logarithmic correction factor \cite{Michon:2023qpl}.

\section{Quantum corrections to strange metal from AdS$_4$/CFT$_3$}\label{sec:AdS4CFT3}

\begin{figure}[htb!]
\begin{center}
\includegraphics[width=15cm, angle=0]{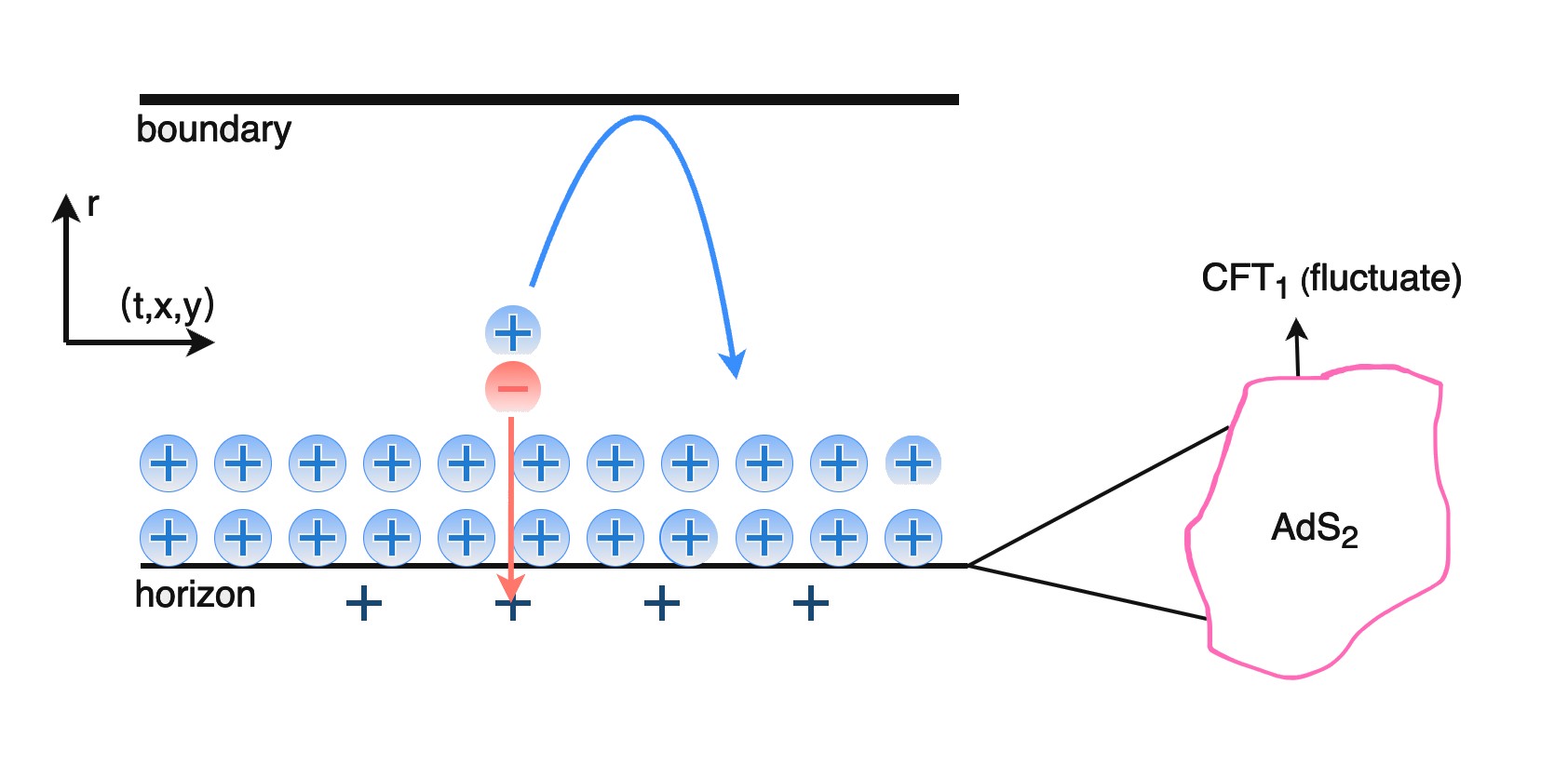}
\caption{ The mechanism of introducing quantum fluctuations in the AdS$_2$ throat region, or equivalently in the boundary CFT$_1$, in contrast to the classical case of \cite{Faulkner:2013bna}.}\label{fig:backgroundfluctuation}
\end{center}
\end{figure}

A pictorial way to describe our work is presented in Fig.~\ref{fig:backgroundfluctuation}. We revisit the holographic framework for transport properties of strange metals developed in \cite{Faulkner:2010da, Faulkner:2009wj, Faulkner:2010zz, Faulkner:2011tm} by including the effects of the strong quantum fluctuations localized in the AdS$_2$ throat region of the bulk geometry. In this section, we discuss how to introduce quantum corrections to the transport properties of strange metal within the framework of AdS$_4$/CFT$_3$. The main idea is first to introduce quantum corrections in the IR CFT$_1$ Green's function \eqref{eq:ads2green}  by considering the fluctuations of the AdS$_2$ or equivalently the boundary CFT$_1$, as shown in Fig.~\ref{fig:backgroundfluctuation}, then lift it to the boundary UV CFT$_3$ Green's function \eqref{eq:ads4green}, and finally obtain the quantum correction to the conductivity \eqref{eq:conductivity} via the Kubo formula.

The limiting near-horizon procedure introduced in Eq.~\eqref{eq:ads2limit} leads to certain zero modes in the gravitational path integral. A careful treatment of these zero modes in near-extremal solutions with an AdS$_2$ throat leads to modifications of the black hole thermodynamics as explained in \cite{Iliesiu:2020qvm, Heydeman:2020hhw, Boruch:2022tno}, and more recently in \cite{Kapec:2023ruw, Rakic:2023vhv} for rotating black holes, including \cite{Maulik:2024dwq} which established the result in various dimensions and extended it to asymptotically AdS black holes. The realization that temperature effectively acts as a coupling constant whereby the high-temperature regime is classical while the very low-temperature regime is quantum and strongly coupled was understood first in the context of two-dimensional JT gravity \cite{Almheiri:2014cka, Jensen:2016pah, Maldacena:2016upp}  (see \cite{Mertens:2022irh, Turiaci:2023wrh} for reviews). The state-of-the-art consists in that any higher-dimensional gravity theory containing near-extremal solutions will admit such zero modes and, consequently, the low-temperature thermodynamics and possibly other dynamical properties will be accordingly corrected.

Let us recall some aspects of a particular dilaton gravity in AdS$_2$, the   JT gravity, which shares some important low-energy symmetry-breaking features with the (0+1)-dimensional quantum-mechanical Sachdev-Ye-Kitaev (SYK) model \cite{Sachdev:1992fk, Kitaev, Maldacena:2016hyu}. After choosing suitable boundary conditions, the fluctuations of the AdS$_2$ spacetime are described by a (0+1)-dimensional Schwarzian action \cite{Maldacena:2016upp}:
\begin{equation}
S[f]=-C\int^{\beta}_0 d\tau \, \{f,\tau\},\quad \text{with } \{f,\tau\} \equiv \frac{f'''}{f'} - \frac{3}{2} \left(\frac{f''}{f'}\right)^2\, ,
\end{equation}
where prime denotes derivative with respect to the Euclidean time $\tau$, and $C$ is the Schwarzian coupling constant with length dimension. One can reach the semi-classical regime by taking $C/\beta$ to be large. The factor $f(\tau)$ is the time reparameterization, and one can introduce time fluctuations by setting $f(\tau)=\tau+\epsilon(\tau)$, where $\epsilon(\tau)$ denotes the boundary time fluctuations. From the Schwarzian action, we can compute the correlation function of the fluctuation $\epsilon$ as \cite{Maldacena:2016upp, Maldacena:2016hyu, Mertens:2022irh}:
\be
  \langle\epsilon(\tau)\, \epsilon(0)\rangle = \frac{1}{2\pi C} \left(\frac{\beta}{2\pi}\right)^3 \left[1-\frac{1}{2}(\tau-\pi)^2+\frac{\pi^2}{6} + \frac{5}{2}{\rm cos}\tau + \left(\frac{\beta}{2\pi}\tau-\pi\right)\, {\rm sin}\tau\right]\, .
\ee

For the Reissner-Nordstr\"om black brane \eqref{eq:metricads4} considered in this paper, we should also take into account the quantum correction from gauge field fluctuations. The effective action for both gravity and gauge fluctuations is given by \cite{Sachdev:2019bjn, Mertens:2019tcm}:
\be\label{eq:effectiveaction}
  S_{eff}[f,\Lambda] = - C\int^{\beta}_{0} d\tau\, \Bigg\{{\rm tan}\frac{\pi}{\beta}f(\tau),\tau\Bigg\} -\frac{K}{2}\int^{\beta}_0 d\tau\, \left[\Lambda'(\tau)-i\mu f'(\tau)\right]^2\, ,
\ee
where $\Lambda(\tau) \equiv \int^{\infty}_{r_0}A_r(r,\tau)\, dr$ denotes the gauge fluctuation, and the coupling constant $K$ is the compressibility of the boundary quantum system, whose value has been discussed in \cite{Davison:2016ngz, Mertens:2019tcm} and is of the same order as $C$.  As shown in Eq.~\eqref{eq:metricads2}, the near-horizon geometry of the Reissner-Nordstr\"om black brane is AdS$_2\times \mathbb{R}^2$. In principle, we should also consider the quantum fluctuations of $\mathbb{R}^2$. These can be dealt with by imposing periodic boundary conditions on $\mathbb{R}^2$, which essentially replaces $\mathbb{R}^2$ by a torus $T^2$ with $U(1) \times U(1)$ isometry, and then performing a Kaluza-Klein reduction. Similar to the AdS$_4$ black hole case \cite{Iliesiu:2020qvm}, the effective action of the JT gravity from the AdS$_4$ black brane should include a contribution from the $U(1) \times U(1)$ gauge field. A similar analysis to the one in \cite{Iliesiu:2020qvm} shows that the energy scale $M_{U(1)\times U(1)}$, at which the $U(1)\times U(1)$ gauge fluctuations start to dominate, satisfies
\be
  M_{U(1)\times U(1)} \ll C^{-1} \simeq K^{-1}\, .
\ee
In this section, we focus on the temperature regime $T \gg M_{U(1)\times U(1)}$. Hence, in \eqref{eq:effectiveaction}, we neglect the quantum fluctuations from $T^2$ and consider it a classical background.

At the classical level, the fermionic operator Green's function in CFT$_1$ can be expressed as
\begin{equation}\label{eq:saddle}
\mathcal{G}(\tau_1,\tau_2)=e^{-\frac{2\pi e_dq}{\beta}(\tau_1-\tau_2)}\left(\frac{1}{\frac{\beta}{\pi}\, {\rm sin}\frac{\pi}{\beta}\, |\tau_1-\tau_2|}\right)^{2\ell}\, .
\end{equation}
By introducing quantum fluctuations, we can transform $\mathcal{G}(\tau_1,\tau_2)$ to an arbitrary solution
\be
  \mathcal{G}'(\tau_1,\tau_2) = e^{-\frac{2\pi e_dq}{\beta}(f(\tau_1)-f(\tau_2))}\, e^{i(\Lambda(\tau_1)-
  \Lambda(\tau_2))}\cdot \left(\frac{\sqrt{f'(\tau_1)f'(\tau_2)}}{\frac{\beta}{\pi}\, {\rm sin}\frac{\pi}{\beta}\, |f(\tau_1)-f(\tau_2)|}\right)^{2\ell}\, .\label{eq:arbitrary}
\ee
Summing all the fluctuations, we obtain the quantum-corrected Green's function as a path integral with the effective action \eqref{eq:effectiveaction}:
\begin{equation}\label{eq:path}
\langle \mathcal{G}(\tau_1,\tau_2)\rangle=\int[\mathcal{D}f][\mathcal{D}\Lambda]\, e^{-S_{eff}[f,\Lambda]}\, \mathcal{G}'(\tau_1,\tau_2)\, .
\end{equation}
It will be convenient to introduce $\tilde{\Lambda}(\tau)=\Lambda(\tau)-i\mu f(\tau)$ which leads to a decoupling of the  $f$-dependent and the $\Lambda$-dependent factors. The path integral \eqref{eq:path} results in 
\begin{align}
\langle \mathcal{G} (\tau_1,\tau_2)\rangle&=\langle e^{i(\tilde{\Lambda}(\tau_1)-\tilde{\Lambda}(\tau_2))}\rangle\cdot \Bigg\langle \left(\frac{\sqrt{f'(\tau_1)f'(\tau_2)}}{\frac{\beta}{\pi}\, {\rm sin}\frac{\pi}{\beta}\, |f(\tau_1)-f(\tau_2)|}\right)^{2\ell}\Bigg\rangle\nonumber\\
& = \langle \mathcal{G}_f\rangle\cdot \langle \mathcal{G}_{\tilde{\Lambda}}\rangle\, ,
\end{align}
where the ${\rm SL}(2,\mathbb{R})$ part, $\langle\mathcal{G}_f\rangle$, and the ${\rm U}(1)$ part, $\langle \mathcal{G}_{\tilde{\Lambda}}\rangle$, have been solved in \cite{Mertens:2017mtv} and \cite{Mertens:2019tcm}, respectively. They have the following explicit expressions:
\begin{align}
\langle \mathcal{G}_{f}\rangle & = \frac{\left(\frac{\beta}{2\pi^2C}\right)^{3/2}}{2\sqrt{\pi}e^{\frac{2\pi^2C}{\beta}}(2C)^{2\ell}}\cdot \int d\zeta(k_1)\, d\zeta(k_2)\, e^{-\mid\tau\mid \frac{k^2_1}{2C} - (\beta-\mid\tau\mid)\frac{k^2_2}{2C}}\, \frac{\Gamma(\ell\pm i(k_1\pm k_2))}{\Gamma(2\ell)}\, ,\label{eq:correction1}\\
\langle \mathcal{G}_{\tilde{\Lambda}}\rangle & = e^{-\frac{\tau(\beta-\tau)}{2K\beta}}e^{\mu\tau}\frac{\theta_3 \left(i\frac{2\pi K}{\beta},-\frac{2\pi K}{\beta}\frac{\mu\beta}{2\pi}-\frac{\tau}{\beta}\right)}{\theta_3 \left(i\frac{2\pi K}{\beta},-\frac{2\pi K}{\beta}\frac{\mu\beta}{2\pi}\right)} \label{eq:correction2}\, ,
\end{align}
where  the chemical potential is denoted by $\mu\equiv - 2\pi e_d q / \beta$.   We have used the standard convention to denote by one $\Gamma$ function in \eqref{eq:correction1} the product of four $\Gamma$ functions with arguments corresponding to all the four sign combinations. 

Our main objective is to incorporate these 
quantum-corrected two-point functions in the computation of some transport properties in the framework of the AdS/CFT applied to the physics of strange metals. We emphasize that we are proceeding in a quasi-phenomenological fashion. Namely, we do not derive the two-point function in the full four-dimensional path integral framework. Instead, we incorporate the corrections to the physics in the throat. We expect, nevertheless, that in the spherically symmetric situations we are discussing here, the approach is completely justified as most other modes decouple in the throat region.

\subsection{DC conductivity}

Having the quantum-corrected CFT$_1$ Green's function, it is natural to explore how they correct transport properties computed in the AdS/CFT framework. We will focus on the DC resistivity as well as the optical conductivity.

In the high-temperature limit ($\beta \to 0$), the quantum fluctuations can be neglected compared to the thermal fluctuations. Hence, in the high-temperature limit, we should recover the result \eqref{eq:dc1} with no quantum fluctuations from gravity.  More appropriately, what we refer to as the high-temperature limit here is simply the limit where quantum fluctuations in the throat are suppressed, which also corresponds to considering the classical geometry of AdS$_2$ as assumed in previous studies of transport \cite{Faulkner:2010da, Faulkner:2009wj, Faulkner:2010zz, Faulkner:2011tm}. We now proceed to verify that intuition as a sanity check.

The quantum fluctuations become dominant only at low temperatures. However, in practice, we would like to express the quantum-corrected result as the known leading-order result with some deviations given by the quantum-correction terms. Hence, we first take the high-temperature limit ($\beta \to 0$) and then gradually lower the temperature by adding terms of higher orders in $\beta$ to see how the resistivity deviates from the linear scaling law. This is an intrinsically perturbative approach driven by the phenomenology of the problem.

For the quantum correction from gravity fluctuations, we start with \eqref{eq:correction1}, which takes the form
\begin{align}
\langle \mathcal{G}_{f}\rangle&
 = \frac{2(\frac{\beta}{2\pi^2C})^{3/2}}{\sqrt{\pi}e^{\frac{2\pi^2C}{\beta}}(2C)^{2\ell}}\int k_1 dk_1\, k_2 dk_2\, {\rm sinh}(2\pi k_1)\, {\rm sinh}(2\pi k_2)\cdot e^{-it \frac{k^2_1}{2C}-(\beta-it)\frac{k^2_2}{2C}}\cdot \frac{1}{\Gamma(2\ell)}\nonumber\\
 & \quad\times \Gamma(\ell+i(k_1+k_2))\cdot \Gamma(\ell-i(k_1+k_2))\cdot \Gamma(\ell+i(k_1-k_2))\cdot \Gamma(\ell-i(k_1-k_2))\, ,\label{eq:g1}
\end{align}
where we  used $d\zeta(k)=dk^2{\rm sinh}(2\pi k)$, and we also assumed $\tau > 0$ and then performed the Wick rotation $\tau\to it$. Following \cite{Lam:2018pvp}, we express $k_{1, 2}$ as
\begin{equation}
k^2_1=2CE_1=2C(M+\omega),\qquad k^2_2=2CE_2=2CM\, ,
\end{equation}
where $E$ and $M$ are the black hole's energy and mass, respectively, and $\omega$ denotes the energy of the particle scattered off the black hole. Consequently, we can rewrite the expression \eqref{eq:g1} as integrals over $M$ and $\omega$. As shown in \cite{Lam:2018pvp}, the integral over $M$ is dominated by $M_0 = 2 \pi^2 C / \beta^2$ in the saddle point approximation justified in the semi-classical limit for large $C$.  We are left with two choices in terms of the remaining parameters, and we now discuss them systematically:

\begin{enumerate}
\item $\omega \ll M$:

This limit implies $C \omega \ll (C T)^2$ and $C E_1 \approx C M\, \propto\, (C T)^2$. Some physical quantities (e.g., DC conductivity) compatible with $\omega \to 0$ and small but finite $T$ can only be defined in this limit. Under this limit, we can still distinguish two sub-cases:
\begin{center}
  (i) $C T \gg 1$; $\quad$ (ii) $C T \ll 1$,
\end{center}
where $C^{-1}$ provides an energy scale when quantum fluctuations start becoming dominant. As mentioned before, in practice, we first take the high-temperature limit ($\beta \to 0$, or equivalently, $C T \gg 1$) and gradually lower the temperature. This is relevant to the holographic strange metal because we do not expect the temperature of a strange metal phase to approach zero. From the phase diagram Fig.~\ref{fig:PhaseDiagram}, we see that the strange metal phase can have a transition to the superconducting phase at a very low temperature, indicating that the holographic model is beyond its regime of validity. Although irrelevant for the physics of interests, we can still take the other limit $C T \ll 1$ for completeness from a mathematical perspective, which is done in App.~\ref{app:the limit CT << 1}.  As expected, the model behaves qualitatively differently in that range of parameters and cannot describe the strange metal phase.

\item $\omega \gg M$:

This limit implies $C \omega \gg (C T)^2$ and $C E_1 \approx C \omega$. Some other physical quantities (e.g., optical conductivity) defined with $T \to 0$ and small but finite $\omega$ can only be defined in this regime of parameters. Similarly, we also distinguish two sub-cases under this limit:
\begin{center}
  (iii) $C \omega \gg 1$; $\quad$ (iv) $C \omega \ll 1$,
\end{center}
where $C^{-1}$ is again the energy scale of the quantum fluctuations. It will become clear in the next subsection that we do not need to take these two limits when computing the optical conductivity for strange metals. However, for mathematical completeness, we also include some expressions in these two limits separately in App.~\ref{app:optical conductivity}.
\end{enumerate}
In this and the next subsections, we will discuss in the framework of quantum-corrected holographic strange metal the DC and the optical conductivities, which are well-defined in the limits $\omega \ll M$ and $\omega \gg M$, respectively.

Let us first consider the DC conductivity, which is compatible with the limit $\omega\ll M$. In this limit, we have
\begin{align}
&k_1\simeq\sqrt{2CM}+\frac{\omega}{2}\sqrt{\frac{2C}{M}},\quad k_2=\sqrt{2CM}\nonumber\\
&\Rightarrow\quad dk_1dk_2=\frac{1}{4M}(2C)\, dM d\omega\, .
\end{align}
By taking $k_1\geq0$ and $k_2\geq0$ and dropping the terms proportional to $\omega/M$, the integral \eqref{eq:g1} can be approximated as
\begin{align}
\langle \mathcal{G}_f(t)\rangle & \approx \frac{\left(\frac{\beta}{2\pi^2C}\right)^{3/2}}{\sqrt{\pi}e^{\frac{2\pi^2C}{\beta}}(2C)^{2\ell-2}}\int \frac{1}{16}dMd\omega e^{4\pi\sqrt{2CM}}e^{-i\omega t-M\beta}\nonumber\\
& \quad\times \frac{1}{\Gamma(2\ell)}\, \Gamma \left(\ell+i(2\sqrt{2CM})\right)\, \Gamma \left(\ell-i(2\sqrt{2CM})\right)\nonumber\\
& \quad\times \Gamma \left(\ell+i\left(\frac{\omega}{2}\sqrt{\frac{2C}{M}}\right)\right)\, \Gamma \left(\ell-i\left(\frac{\omega}{2}\sqrt{\frac{2C}{M}}\right)\right)\, .
\end{align}
Since $\ell$ is of the same order as the particle mass, i.e., $\ell\ll M$, the first two Gamma functions can be approximately computed as
\begin{align}
  \Gamma(\ell+i2\sqrt{2CM})\, \Gamma(\ell-i2\sqrt{2CM})\approx \frac{\pi2\sqrt{2CM}}{{\rm sinh}\pi2\sqrt{2CM}}(8CM)^{\ell-1}\approx 2\frac{\pi(2\sqrt{2CM})^{2\ell-1}}{e^{2\pi\sqrt{2CM}}}\, ,
\end{align}
where we used the identity
\be
  \Gamma(\ell+i2\sqrt{2CM})\, \Gamma(\ell-i2\sqrt{2CM})=\frac{\pi2\sqrt{2CM}}{{\rm sinh}\pi2\sqrt{2CM}}\prod^{\ell-1}_{n=1}(n^2+8CM)\, .
\ee
Plugging the above-approximated expression back and performing a Fourier transform, we obtain
\begin{align}
\langle \mathcal{G}_f(\omega)\rangle & = \frac{\sqrt{\pi}\left(\frac{\beta}{2\pi^2C}\right)^{3/2}}{4\sqrt{\pi}e^{\frac{2\pi^2C}{\beta}}(2C)^{2\ell-2}}\frac{1}{\Gamma(2\ell)}\int dM e^{2\pi\sqrt{2CM}-M\beta}\nonumber\\
& \quad \times \Gamma\left(\ell+i\frac{\omega}{2}\sqrt{\frac{2C}{M}}\right)\Gamma\left(\ell-i\frac{\omega}{2}\sqrt{\frac{2C}{M}}\right)(2\sqrt{2CM})^{2\ell-1}\, .\label{eq:g2}
\end{align}
We evaluate this integral using a saddle-point approximation, with the final result given by:

\begin{align}
{} & \langle \mathcal{G}_f(\omega)\rangle\nonumber\\
 \approx\,\, & \frac{\sqrt{\pi} \left(\frac{\beta}{2\pi^2C}\right)^{3/2}}{4(2C)^{2\ell-2}} \frac{(2\pi)^{2\ell-1}}{\Gamma(2\ell)} \left[\frac{\beta}{2C}+\frac{1-2\ell}{2\pi^2} \left(\frac{\beta}{2C}\right)^2\right]^{1-2\ell}\cdot \frac{\pi}{{\rm sin} \left(\pi\ell + iC\omega \left[\frac{\beta}{2C}+\frac{(1-2\ell)}{2\pi^2} \left(\frac{\beta}{2C}\right)^2\right]\right)}\nonumber\\
 & \times \frac{\Gamma \left(\ell - i\frac{C\omega}{\pi} \left[\frac{\beta}{2C}+\frac{(1-2\ell)}{2\pi^2} \left(\frac{\beta}{2C}\right)^2\right]\right)}{\Gamma \left(1 - \ell - i\frac{C\omega}{\pi} \left[\frac{\beta}{2C}+\frac{(1-2\ell)}{2\pi^2} \left(\frac{\beta}{2C}\right)^2\right]\right)}\, .\label{eq:c1}
\end{align}
The derivation follows fairly standard steps, which we describe in detail in App.~\ref{Sec:AppendixA}.

Compared to the quantum-corrected results, the previous results in the literature correspond to the high-temperature limit defined by $T > C^{-1}$, where quantum fluctuations are negligible compared to thermal fluctuations. Therefore, to clearly see the quantum corrections to the previous results, we have to expand the new quantum-corrected results in the high-temperature limit (i.e., small $\beta$).

Comparing the quantum-corrected Green's function \eqref{eq:c1} and the tree-level Green's function \eqref{eq:ads2green}, we interpret the terms of higher orders in $\beta$ as the quantum correction from gravity fluctuations. Note that we keep terms to the order $\beta^2$. One can check that when taking the high-temperature limit and dropping the terms $\sim \beta^2$, Eq.~\eqref{eq:c1} above goes back to Eq.~(4.10) in \cite{Lam:2018pvp}, i.e., the classical Green's function.

Let us move on to the quantum correction from gauge fluctuations. We start with $\langle \mathcal{G}_{\tilde{\Lambda}}\rangle$ defined in \eqref{eq:correction2}. In appendix \ref{Sec:AppendixA}, we show the explicit details of how one can use properties of the theta function, $\theta_3$, to approximate the expression in the $\beta \to 0$ regime. After performing a Wick rotation and doing a Fourier transform,  one obtains, in the  semi-classical limit (i.e., $K \to \infty$):
\begin{align}
\langle \mathcal{G}_{\tilde{\Lambda}}(\omega)\rangle & = \delta(\omega+\mu) - \sum_{m=1}e^{-\pi \frac{2\pi K}{\beta}m^2} e^{-K (i\,\mu\, 2\pi m)}\delta(\omega+\mu)\nonumber\\
{} & \quad + \sum_{m=1}e^{-\pi \frac{2\pi K}{\beta}m^2} e^{-K (i\,\mu\, 2\pi m)} \delta \left(\omega+\mu-\frac{i(2\pi m)}{\beta}\right)\nonumber\\
{} & \quad + \sum_{m=1}e^{-\pi \frac{2\pi K}{\beta}m^2} e^{-K (i\,\mu\, 2\pi m)} \delta \left(\omega+\mu-\frac{i(2\pi m)}{\beta}\right)\cdot \sum_{n=1} e^{-\pi \frac{2\pi K}{\beta}n^2} e^{-K (i\,\mu\, 2\pi n)}\, ,\label{eq:gaugecorrection1'}
\end{align}
where $\mu \equiv -\frac{2\pi e_d q}{\beta}$. Since $K$ and $1/\beta$ are large, the exponential factors are highly suppressed for $m > 1$. Thus, we can keep only the dominant terms with $m=1$ as the leading-order approximation, and consequently, Eq.~\eqref{eq:gaugecorrection1'} becomes
\begin{align}
\langle \mathcal{G}_{\tilde{\Lambda}}(\omega)\rangle & = \delta(\omega+\mu) - e^{-\pi\frac{2\pi K}{\beta}} e^{-i\, 2\pi\mu K}\delta(\omega+\mu)\nonumber\\
{} & \quad + (e^{-\pi\frac{2\pi K}{\beta}} e^{-i\, 2\pi\mu K} + e^{-\pi\frac{4\pi K}{\beta}}e^{-i\, 4\pi\mu K})\, \delta \left(\omega + \mu - \frac{i\, 2\pi}{\beta}\right)\, .\label{eq:c2}
\end{align}

Combining the quantum-corrected Green's function \eqref{eq:c1} from gravity fluctuations and the quantum-corrected Green's function \eqref{eq:c2} from gauge fluctuations, we have an explicit expression for the complete quantum-corrected IR CFT$_1$ Green's function. To compare it with the tree-level Green's function \eqref{eq:ads2green}, we need to introduce an additional constant factor
\begin{equation}
(-1)^{\ell}\, 2^{2\ell-1}\, \frac{\Gamma(-2\ell+1)}{\Gamma(2\ell-1)}\frac{\Gamma(\ell+\frac{1}{2}-iqe_d)}{\Gamma(\frac{3}{2}-\ell-iqe_d)}\cdot\frac{(m-i\tilde{m})R_2-iqe_d-\ell+\frac{1}{2}}{(m-i\tilde{m})R_2-iqe_d+\ell-\frac{1}{2}}\, ,
\end{equation}
which is independent of $\omega$ and $T$. Consequently, the Green's function with full quantum corrections becomes
\begin{align}
\langle \mathcal{G}(\omega)\rangle & = (-1)^{\ell}2^{2\ell-1}\frac{\Gamma(-2\ell+1)}{\Gamma(2\ell-1)}\frac{\Gamma(\ell+\frac{1}{2}-iq e_d)}{\Gamma(\frac{3}{2}-\ell-iqe_d)}\cdot\frac{(m-i\tilde{m})R_2 - iq e_d-\ell+\frac{1}{2}}{(m-i\tilde{m}) R_2-i q e_d+\ell-\frac{1}{2}}\nonumber\\
& \,\,\times \Bigg[(1-e^{-\pi\frac{2\pi K}{\beta}}e^{-i2\pi\mu K})\frac{\sqrt{\pi} \left(\frac{\beta}{2\pi^2C}\right)^{3/2}}{4(2C)^{2\ell-2}}\frac{1}{\Gamma(2\ell)} (2\pi)^{2\ell-1} \left[\frac{\beta}{2C}+\frac{1-2\ell}{2\pi^2} \left(\frac{\beta}{2C}\right)^2\right]^{1-2\ell}\nonumber\\
& \qquad\times \frac{\pi}{{\rm sin}(\pi\ell+iC(\omega-\frac{2\pi e_d q}{\beta})[\frac{\beta}{2C}+\frac{(1-2\ell)}{2\pi^2}(\frac{\beta}{2C})^2])}\frac{\Gamma(\ell-i\frac{C}{\pi}(\omega-\frac{2\pi e_d q}{\beta})[\frac{\beta}{2C}+\frac{(1-2\ell)}{2\pi^2}(\frac{\beta}{2C})^2])}{\Gamma(1-\ell-i\frac{C}{\pi}(\omega-\frac{2\pi e_d q}{\beta})[\frac{\beta}{2C}+\frac{(1-2\ell)}{2\pi^2}(\frac{\beta}{2C})^2])}\nonumber\\
& \quad  +\left(e^{-\pi\frac{2\pi K}{\beta}}e^{-i2\pi\mu K}+e^{-\pi\frac{4\pi K}{\beta}}e^{-i4\pi\mu K}\right)\frac{\sqrt{\pi} \left(\frac{\beta}{2\pi^2C}\right)^{3/2}}{4(2C)^{2\ell-2}}\frac{1}{\Gamma(2\ell)} (2\pi)^{2\ell-1} \nonumber\\
& \qquad\times \left[\frac{\beta}{2C}+\frac{1-2\ell}{2\pi^2} \left(\frac{\beta}{2C}\right)^2\right]^{1-2\ell}\frac{\pi}{{\rm sin} (\pi\ell+iC(\omega-\frac{2\pi e_d q}{\beta}-\frac{i2\pi}{\beta})[\frac{\beta}{2C} + \frac{(1-2\ell)}{2\pi^2}(\frac{\beta}{2C})^2])}\nonumber\\
& \qquad\times \frac{\Gamma \left(\ell-i\frac{C}{\pi}(\omega-\frac{2\pi e_d q}{\beta}-\frac{i2\pi}{\beta})\,\, \left[\frac{\beta}{2C}+\frac{(1-2\ell)}{2\pi^2}(\frac{\beta}{2C})^2\right] \right)}{\Gamma \left(1-\ell-i\frac{C}{\pi}(\omega-\frac{2\pi e_d q}{\beta}-\frac{i2\pi}{\beta})\,\, \left[\frac{\beta}{2C}+\frac{(1-2\ell)}{2\pi^2}(\frac{\beta}{2C})^2\right] \right)}\Bigg]\, .\label{eq:complete1}
\end{align}

\begin{figure}[htb!]
\begin{center}
\includegraphics[width=10.5cm, angle=0]{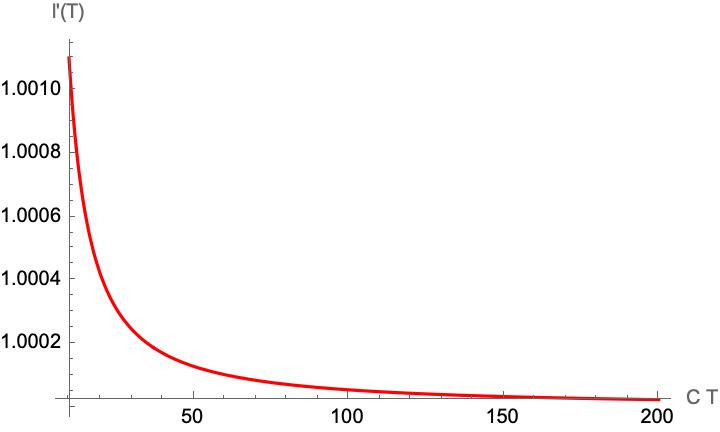}
\caption{The renormalized parameter $\ell'$ as a function of $C T$ (with a special choice of parameters $\ell=1$, $q=1$, $K=10^4$, and $\omega=10^{-3}$)}\label{fig:dcc1}
\end{center}
\end{figure}

\begin{figure}[htb!]
\begin{center}
\includegraphics[width=10cm, angle=0]{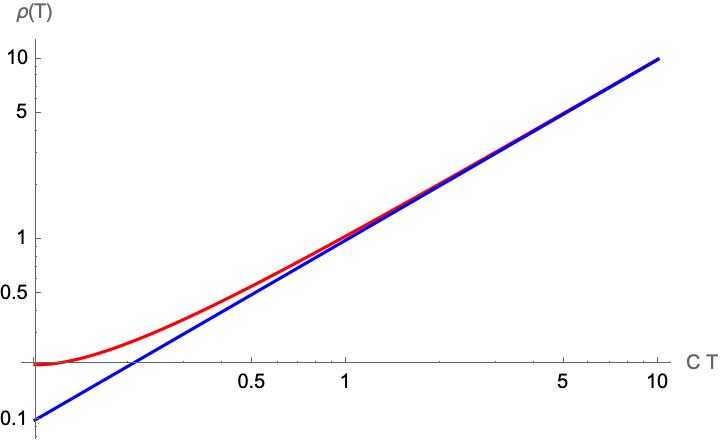}
\caption{The uncorrected (blue) and quantum-corrected (red) DC resistivities as functions of $C T$ (with a special choice of parameters $\ell=1$, $q=1$, $K=10^4$, and $\omega=10^{-3}$)}\label{fig:dcc2}
\end{center}
\end{figure}

To see how quantum corrections affect the DC conductivity of strange metal, let us rewrite the tree-level Green's function \eqref{eq:ads2green} and the quantum-corrected Green's function \eqref{eq:complete1} in similar forms:
\begin{align}
\mathcal{G}(\omega) & = F_1(\beta,\ell;\omega,q,K)\cdot \left(\frac{\beta}{2C}\right)^{1-2\ell}\, ,\nonumber\\
\langle \mathcal{G}(\omega)\rangle & = F_2(\beta,\ell;\omega,q,K)\cdot \left[\frac{\beta}{2C}+\frac{1-2\ell}{2\pi^2}\left(\frac{\beta}{2C}\right)^2\right]^{1-2\ell}\, ,
\end{align}
where the two factors $F_1$ and $F_2$ are defined as follows:
\begin{align}
F_1(\beta,\ell;\omega,q,K) & = (4\pi)^{2\ell-1}(2C)^{1-2\ell}\frac{\Gamma(-2\ell+1)}{\Gamma(2\ell-1)}\frac{\Gamma(\ell+\frac{1}{2}-iqe_d)}{\Gamma(\frac{3}{2}-\ell-iqe_d)}\cdot \frac{(m-i\tilde{m}) R_2 - iq e_d-\ell+\frac{1}{2}}{(m-i\tilde{m}) R_2 - iq e_d+\ell-\frac{1}{2}}\nonumber\\
{} & \quad\times \frac{\Gamma(\ell-\frac{i\omega}{2\pi T} + iq e_d)}{\Gamma(1-\ell-\frac{i\omega}{2\pi T} + iq e_d)}\, ,
\end{align}
\begin{align}
F_2(\beta,\ell;\omega,q,K) & = (-1)^{\ell}2^{2\ell-1}\frac{\Gamma(-2\ell+1)}{\Gamma(2\ell-1)}\frac{\Gamma(\ell+\frac{1}{2}-iqe_d)}{\Gamma(\frac{3}{2}-\ell-iqe_d)}\cdot\frac{(m-i\tilde{m})R_2-iqe_d-\ell+\frac{1}{2}}{(m-i\tilde{m})R_2-iqe_d+\ell-\frac{1}{2}}\nonumber\\
{} & \quad\times \Bigg[(1-e^{-\pi\frac{2\pi K}{\beta}}e^{-i2\pi\mu K})\frac{\sqrt{\pi} \left(\frac{\beta}{2\pi^2C}\right)^{3/2}}{4(2C)^{2\ell-2}}\frac{1}{\Gamma(2\ell)} (2\pi)^{2\ell-1}\nonumber\\
{} & \qquad\quad\times \frac{\pi}{{\rm sin}(\pi\ell+iC(\omega-\frac{2\pi e_dq}{\beta}) [\frac{\beta}{2C}+\frac{(1-2\ell)}{2\pi^2}(\frac{\beta}{2C})^2])}\nonumber\\
{} & \qquad\quad\times \frac{\Gamma(\ell-i\frac{C}{\pi}(\omega-\frac{2\pi e_dq}{\beta})[\frac{\beta}{2C}+\frac{(1-2\ell)}{2\pi^2}(\frac{\beta}{2C})^2])}{\Gamma(1-\ell-i\frac{C}{\pi}(\omega-\frac{2\pi e_dq}{\beta})[\frac{\beta}{2C}+\frac{(1-2\ell)}{2\pi^2}(\frac{\beta}{2C})^2])}\nonumber\\
{} & \qquad +(e^{-\pi\frac{2\pi K}{\beta}}e^{-i2\pi\mu K}+e^{-\pi\frac{4\pi K}{\beta}}e^{-i4\pi\mu K})\frac{\sqrt{\pi} \left(\frac{\beta}{2\pi^2C}\right)^{3/2}}{4(2C)^{2\ell-2}}\frac{1}{\Gamma(2\ell)} (2\pi)^{2\ell-1}\nonumber\\
{} & \qquad\quad\times \frac{\pi}{{\rm sin}(\pi\ell+iC(\omega-\frac{2\pi e_dq}{\beta}-\frac{i2\pi}{\beta})[\frac{\beta}{2C} + \frac{(1-2\ell)}{2\pi^2}(\frac{\beta}{2C})^2])}\nonumber\\
{} & \qquad\quad\times \frac{\Gamma(\ell-i\frac{C}{\pi}(\omega-\frac{2\pi e_dq}{\beta}-\frac{i2\pi}{\beta})[\frac{\beta}{2C} + \frac{(1-2\ell)}{2\pi^2}(\frac{\beta}{2C})^2])}{\Gamma(1-\ell-i\frac{C}{\pi}(\omega-\frac{2\pi e_dq}{\beta}-\frac{i2\pi}{\beta})[\frac{\beta}{2C}+\frac{(1-2\ell)}{2\pi^2}(\frac{\beta}{2C})^2])}\Bigg]\, .
\end{align}
In the spirit of renormalization, we can rewrite the quantum-corrected Green's function in the same form as the tree-level Green's function:
\begin{align}
\langle \mathcal{G} (\omega, T)\rangle & = F_2(\beta,\ell;\omega,q,K)\cdot \left[\frac{\beta}{2C} + \frac{1-2\ell}{2\pi^2} \left(\frac{\beta}{2C}\right)^2\right]^{1-2\ell}\nonumber\\
 & = F_1(\beta,\ell;\omega,q,K)\cdot \left(\frac{\beta}{2C}\right)^{1-2\ell'}\, ,\label{eq:Green Fct with F1}
\end{align}
where we have defined a renormalized parameter $\ell'$ to represent the quantum-corrected $\ell$. This renormalized $\ell'$ has the following explicit expression:
\begin{equation}
\ell'=\frac{1}{2}-\frac{{\rm ln} \left(\frac{F_2(\beta,\ell;\omega,q,K)}{F_1(\beta,\ell;\omega,q,K)}\right) + (1-2\ell)\, {\rm ln} \left[\frac{\beta}{2C}+\frac{1-2\ell}{2\pi^2} \left(\frac{\beta}{2C}\right)^2\right]}{2\, {\rm ln}\, \frac{\beta}{2C}}\, ,
\end{equation}
which is also shown in Fig.~\ref{fig:dcc1} as a function of $C\,  T$. The deviation of the renormalized parameter $\ell'$ from the classical value $\ell = 1$ manifests for small $C\, T$.
Because the quantum-corrected Green's function with the renormalized parameter $\ell'$ has the same form as the tree-level Green's function, we can lift the quantum-corrected IR CFT$_1$ Green's function $\langle \mathcal{G} (\omega, T)\rangle$ \eqref{eq:Green Fct with F1} to the UV CFT$_3$ Green's function $G_R (\omega, T)$ and perform a calculation similar to the one without quantum corrections \cite{Faulkner:2013bna}. Finally, we obtain the DC resistivity of strange metal, including quantum correction, which at the leading order is
\begin{equation}
\rho\simeq \alpha T^{2\ell'-1}\, .
\end{equation}
The quantum-corrected resistivity $\rho$ is shown in Fig.~\ref{fig:dcc2}. When $T < C^{-1}$, the quantum fluctuations become dominant, and the quantum-corrected resistivity of strange metal clearly deviates from the linear scaling law. The current experimental data for the resistivity of strange metal is $\sim T^{1 \pm 0.05}$ \cite{PhysRevMaterials.2.024804, Nguyen2021SuperconductivityIA}, which is insufficient to distinguish the quantum correction clearly. We expect that more precise experimental data in the future can see some deviations from the strict linear resistivity at relatively low temperatures.

\subsection{Optical conductivity}

The optical conductivity is measured at nearly zero temperature. Hence, we have to consider a different approximation for Eqs.~\eqref{eq:correction1} and \eqref{eq:correction2}, i.e., in the low-temperature limit $\beta\to\infty$ with small but nonzero $\omega$. In this section, we focus on the frequency regime $\omega \gg M_{U(1)\times U(1)}$. Thus, we can neglect the quantum fluctuations from the torus $T^2$ like in the last section.

For $\beta\to\infty$, the quantum-corrected Green's function \eqref{eq:correction1} from gravity fluctuations reduces to
\begin{equation}
\langle\mathcal{G}_f(\tau)\rangle = \frac{1}{(2C)^{2\ell}}\int dk^2{\rm sinh}(2\pi k)\, e^{-\mid\tau\mid \frac{k^2}{2C}}\frac{\Gamma^2(\ell+ik)\, \Gamma^2(\ell-ik)}{2\pi^2\Gamma(2\ell)}\, .
\end{equation}
After setting $k^2=2C\omega$, $k=\sqrt{2C\omega}$, and $dk=\frac{1}{2}\sqrt{2C}\omega^{-1/2}d\omega$, it becomes
\begin{align}
\langle\mathcal{G}_f(\tau)\rangle & = \frac{1}{(2C)^{2\ell}}\int d\omega\, {\rm sinh}(2\pi\sqrt{2C\omega})\, e^{-\mid\tau\mid\omega}\cdot \frac{\Gamma^2(\ell+i\sqrt{2C\omega})\, \Gamma^2(\ell-i\sqrt{2C\omega})}{2\pi^2\, \Gamma(2\ell)}\, .
\end{align}
Assuming $\tau > 0$, performing a Wick rotation, and subsequently doing a Fourier transform, we obtain
\begin{align}
\langle\mathcal{G}_f(\omega)\rangle & = \frac{1}{(2C)^{2\ell}}\, {\rm sinh}(2\pi\sqrt{2C\omega})\, \frac{\Gamma^2(\ell+i\sqrt{2C\omega})\, \Gamma^2(\ell-i\sqrt{2C\omega})}{2\pi^2\, \Gamma(2\ell)}\nonumber\\
{} & = \frac{1}{4\pi^2(2C)^{2\ell}}\, \frac{1}{\Gamma(2\ell)}\, \left(e^{2\pi\sqrt{2C\omega}} - e^{-2\pi\sqrt{2C\omega}}\right)\cdot \Gamma^2(\ell+i\sqrt{2C\omega})\, \Gamma^2(\ell-2\sqrt{2C\omega})\, .\label{eq:c21}
\end{align}

For gauge fluctuations, by taking the low-temperature limit $(\beta\to\infty)$ for Eq.~\eqref{eq:correction2}, we have
\begin{equation}
\langle \mathcal{G}_{\tilde{\Lambda}}\rangle=e^{-\frac{\tau}{2K}+\mu\tau}\, .
\end{equation}
Performing a Fourier transform, this correction due to gauge fluctuations provides a $\delta$-function:
\begin{equation}\label{eq:c22}
  \delta \left(\omega-\frac{1}{2K}+\mu\right)\, .
\end{equation}

Combining two parts of quantum corrections can be done by performing a convolution of \eqref{eq:c21} and \eqref{eq:c22}, which essentially introduces a shift of $\omega$ in \eqref{eq:c21}. After introducing a constant factor to match the tree-level result, we have the full quantum-corrected Green's function:
\begin{align}
\langle \mathcal{G}(\omega)\rangle&=(-1)^{\ell}2^{2\ell-1}\frac{\Gamma(-2\ell+1)}{\Gamma(2\ell-1)}\frac{\Gamma(\ell+\frac{1}{2}-iqe_d)}{\Gamma(\frac{3}{2}-\ell-iqe_d)}\frac{(m-i\tilde{m})R_2-iqe_d-\ell+\frac{1}{2}}{(m-i\tilde{m})R_2-iqe_d+\ell-\frac{1}{2}}\frac{1}{4\pi^2(2C)^{2\ell}}\frac{1}{\Gamma(2\ell)}\nonumber\\
{} & \quad\times \left(e^{2\pi\sqrt{2C(\omega-\frac{1}{2K}+\mu)}} - e^{-2\pi\sqrt{2C(\omega-\frac{1}{2K}+\mu)}}\right)\nonumber\\
& \quad\times \Gamma^2 \left(\ell+i\sqrt{2C \left(\omega-\frac{1}{2K}+\mu\right)}\right)\, \Gamma^2 \left(\ell-2\sqrt{2C \left(\omega-\frac{1}{2K}+\mu\right)}\right)\, .\label{eq:cc2}
\end{align}

We again rewrite the quantum-corrected Green's function in the same form as the tree-level one:
\begin{align}
\mathcal{G}(\omega) & = F_3(\omega,\ell;q,K)\, (2C\omega)^{2\ell-1}\, \\
\langle \mathcal{G}(\omega)\rangle & = F_4(\omega,\ell;q,K) = F_3(\omega,\ell;q,K)\, (2C\omega)^{2\ell'-1}\, ,\label{eq:Green Fct with F3}
\end{align}
where
\begin{equation}
\ell'=\frac{1}{2}+\frac{1}{2}\frac{{\rm ln}\frac{F_4(\omega,\ell;q,K)}{F_3(\omega,\ell;q,K)}}{{\rm ln}(2C\omega)}\, ,
\end{equation}
and
\begin{align}
\frac{F_4(\omega,\ell;q,K)}{F_3(\omega,\ell;q,K)} & = (-1)^{\ell}e^{i\pi(\ell-\frac{1}{2})}\frac{1}{8\pi^2}\frac{1}{\Gamma(2\ell)}\frac{1}{C}\cdot \left(e^{2\pi\sqrt{2C(\omega-\frac{1}{2K}+\mu)}}-e^{-2\pi\sqrt{2C(\omega-\frac{1}{2K}+\mu)}}\right)\nonumber\\
& \quad\times \Gamma^2 \left(\ell+i\sqrt{2C \left(\omega-\frac{1}{2K}+\mu\right)}\right)\cdot \Gamma^2 \left(\ell-2\sqrt{2C \left(\omega-\frac{1}{2K}+\mu\right)}\right)\, .
\end{align}
We notice that $\mu=-2\pi qe_d/\beta\to0$ in the low-temperature limit $\beta\to\infty$.

\begin{figure}[htb!]
\begin{center}
\includegraphics[width=10cm, angle=0]{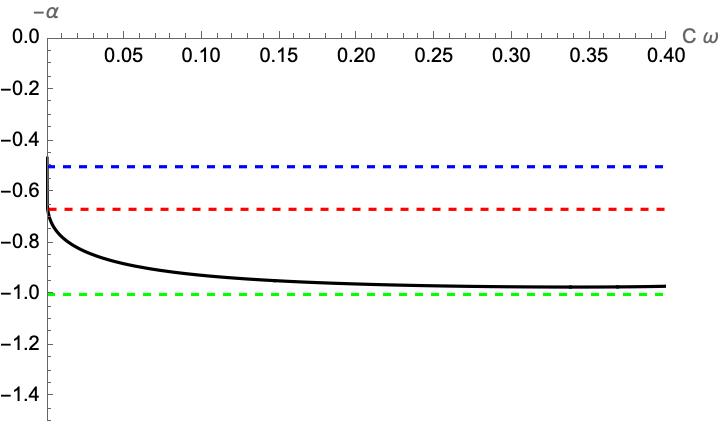}
\caption{The quantum-corrected $-\alpha$ as a function of $C \omega$ (with a special choice of parameters $\ell=1$, $q=1$, $K=10^4$, and $\mu = 0$). The blue, red, and green dashed lines denote the special values $-1/2$, $-2/3$, and $-1$, respectively.}\label{fig:optical}
\end{center}
\end{figure}

Starting from the quantum-corrected IR CFT$_1$ Green's function $\langle \mathcal{G} (\omega) \rangle$ \eqref{eq:Green Fct with F3}, we can again lift it to the UV CFT$_3$ Green's function $G_R (\omega)$ and perform a calculation similar to the one without quantum corrections \cite{Faulkner:2013bna}. Finally, we obtain the quantum-corrected optical conductivity:
\begin{equation}
\sigma\sim\frac{ic}{\omega^{2\ell'-1}} \left(\frac{1}{{\rm log}\omega}+\frac{1}{({\rm log}\omega)^2}\frac{1+i\pi}{2}\right) + \cdots
\end{equation}
The leading-order power of $\omega$ in optical conductivity (i.e., $1 - 2 l'$) is plotted in Fig.~\ref{fig:optical}. We see when $C\omega$ approaches zero, the power can take the values ranging from $-1$ to $0$, which recover all the experimental results for optical conductivity, i.e., the values of $-\alpha$ in $\sigma (\omega) \sim \omega^{-\alpha}$ mentioned in the introduction. Thus, we can fit all the experimental values of $\alpha$: The most typical value is 2/3 \cite{PhysRevB.58.11631, Marel2006ScalingPO, Phillips:2022nxs}, and other experimentally measured values include 0.5 \cite{PhysRevB.66.041104}, 0.6 \cite{vandeMarel:2003wn}, 0.65 \cite{vandeMarel:2003wn}, 0.7 \cite{Baraduc1996InfraredCI}, 0.77 \cite{PhysRevB.49.9846}, and 0.8 \cite{Michon:2023qpl}. Hence, we provide a theoretical explanation for all the observed data. It strongly suggests that the optical conductivity is directly related to the quantum corrections at low temperatures, and different experimental values of $\alpha$ could be interpreted as measurements done at slightly different energies.

\section{Quantum corrections to strange metal from complex SYK model}\label{sec:CplxSYK}

The paradigmatic SYK-like models are (0+1)-d quantum-mechanical theories of fermions with random all-to-all interactions. These models have been considered to describe compressible quantum many-body systems without quasiparticle excitations as examples of non-Fermi liquids. In this class of theories, the complex SYK model \cite{Davison:2016ngz} is a special case, which considers $N$ canonical complex fermions $f_i$ labeled by $i=1,\cdots,N$, whose Hamiltonian is
\begin{align}
  H_0 & = \sum J_{i_1\,i_2\,\cdots\, i_p} f^{\dagger}_{i_1}f^{\dagger}_{i_2}\cdots f^{\dagger}_{i_{p/2}}f_{i_{p/2+1}}\cdots f_{i_{p-1}}f_{i_p},\nonumber\\
  1 \leq i_1 < i_2 < & \cdots < i_{p/2}\leq N,\quad 1\leq i_{p/2+1} < i_{p/2+2} < \cdots < i_p\leq N\, ,
\end{align}
where $p$ is an even integer, and the couplings $J_{i_1\,i_2\,\cdots\, i_p}$ are random complex numbers with zero mean. The complex SYK model can be solved in the large-$N$ limit via the Schwinger-Dyson approach, and the resulting Green's function for complex fermions is:  
\begin{equation}
\mathcal{G}_R(\tau,T)=e^{-\frac{2\pi e_dq}{\beta}\tau}\left(\frac{\pi T}{{\rm sin}(\pi T\tau)}\right)^{2\Delta}\, ,
\end{equation}
where $\Delta=1/p$. This Green's function is the same as spinor saddle-point Green's function \eqref{eq:saddle} in CFT$_1$. We will denote $\Delta$ by $\ell$ what follows. We can also write this saddle-point Green's function in frequency space up to some constant as 
\begin{equation}\label{eq:saddleomega}
\mathcal{G}_R(\omega,T)=(2\pi T)^{2\ell-1}\frac{\Gamma(\ell-\frac{i}{2\pi T}(\omega-2\pi qe_dT))}{\Gamma(1-\ell-\frac{i}{2\pi T}(\omega-2\pi qe_dT))}\, .
\end{equation}

In this section, we review the complex SYK model as an effective description for strange metal closely following the presentation of \cite{Davison:2016ngz}. Previous direct discussions of the SYK model as an explanation for linear resistivity were presented, for example, \cite{Guo:2020aog, Cha_2020}. We note, vis-\'a-vis the holographic approach and JT gravity specifically, that all these models have the same pattern of low energy symmetry breaking describing time reparametrizations and a $U(1)$ gauge field; this is the universality that validates holographic approaches.

\subsection{DC conductivity}

When we consider the complex SYK model, the DC conductivity from the Kubo formula is given by \cite{RevModPhys.94.035004}:
\begin{equation}\label{eq:sykkubo}
\sigma_{{\rm dc}}=\frac{2\pi e^2}{\bar{h}}\int d\omega\frac{\beta}{4{\rm cosh^2}(\frac{\beta\omega}{2})}\int d\epsilon\, \phi(\epsilon) A^2(\epsilon,\omega,T)\, ,
\end{equation}
where $\epsilon$ is the energy of a bare single-particle state, $\phi(\epsilon)$ is a transport function defined on Bravais lattice, and $A(\epsilon,\omega,T) = -(1/\pi)\, {\rm Im}\,\mathcal{G}_R(\epsilon,\omega,T)\, \propto\,{\rm Im}\,\Sigma^{-1}(\epsilon,\omega,T)$ is the energy (momentum) resolved spectral function. The Green's function $\mathcal{G}_R(\epsilon,\omega, T)$ has the same expression as in Eq.~\eqref{eq:saddleomega}, i.e., with the same $(\omega, T)$-dependence. In the low-temperature limit, the imaginary part of the self-energy ${\rm Im} \Sigma$ is much smaller than the dispersion of the band (i.e., the range over which $\epsilon$ varies in the integral). Then, the integral over $\epsilon$ in \eqref{eq:sykkubo} can be approximately written as \cite{RevModPhys.94.035004}:
\be\label{eq:Integral}
  \int d\epsilon\phi(\epsilon)A^2(\epsilon,\omega) \simeq \frac{\phi \left[\omega+\mu-{\rm Re\Sigma(\omega)}\right]}{2\pi\, |{\rm Im}\Sigma(\omega)|}\, .
\ee
Due to the Fermi factor ${\rm cosh}(\beta\omega/2)$ only $\omega\ll T$ is relevant, thus we can set $\omega=0$ in the numerator of \eqref{eq:Integral}. Defining a renormalized Fermi energy $\epsilon_F \equiv \mu - {\rm Re}\Sigma(0,0)$, we obtain the DC conductivity
\begin{equation}
  \sigma_{dc}\simeq\frac{e^2\phi(\epsilon_F)}{\bar{h}}\int d\omega\frac{\beta}{4\, {\rm cosh}^2(\beta\omega/2)}\frac{1}{|{\rm Im}\Sigma(\omega,T)|}\, ,
\end{equation}
where
\be
  {\rm Im}\Sigma(\omega,T)\,\,\propto\,\, \mathcal{G}_{R}^{-1}(\omega,T)=(2\pi T)^{1-2\ell}\frac{\Gamma(1-\ell-\frac{i}{2\pi T}(\omega-2\pi qe_dT))}{\Gamma(\ell-\frac{i}{2\pi T}(\omega-2\pi qe_dT))}\, .
\ee
We can further rescale $\omega$ as $\omega\to\beta^{-1}\omega$ to obtain
\begin{align}
  \sigma_{dc}\,\, & \propto\,\, \int d\omega\, \frac{1}{4{\rm cosh}^2(\frac{\omega}{2})}\, \frac{1}{{\rm Im}\Sigma(\omega,T)}\, ,\nonumber\\
  {\rm Im}\Sigma(\omega,T)\,\, & \propto\,\, (2\pi T)^{1-2\ell}\frac{\Gamma(1-\ell-\frac{i}{2\pi}(\omega-2\pi qe_d))}{\Gamma(\ell-\frac{i}{2\pi}(\omega-2\pi qe_d))}\, .
\end{align}
Finally, the temperature dependence for DC resistivity is
\begin{equation}\label{eq:sykdc}
\rho_{dc}\,\, \propto\,\, T^{1-2\ell}\, .
\end{equation}
We note that this relation is quite different from the temperature dependence \eqref{eq:dc1} of the DC resistivity from the AdS$_4$/CFT$_3$ approach \cite{Faulkner:2013bna, Faulkner:2010zz}. The physical reason is that in the AdS$_4$/CFT$_3$ approach \cite{Faulkner:2013bna, Faulkner:2010zz} the holographic strange metal is defined on a (2+1)-dimensional boundary in the UV, while the (0+1)-dimensional complex SYK model \cite{Davison:2016ngz} is defined in the IR, corresponding to the near-horizon region of the AdS$_4$ black brane.

Now, in the (0+1)-dimensional complex SYK model description, the linear resistivity of strange metal can be achieved by taking $\ell=0$ in \eqref{eq:sykdc}, which corresponds to $p\to\infty$ in complex SYK model.

The next step is to consider quantum correction in the complex SYK model. The quantum-corrected Green's function in the high-temperature limit has the same expression as in Eq.~\eqref{eq:Green Fct with F1} since the tree-level complex fermion Green's function \eqref{eq:saddleomega} has the same form as \eqref{eq:saddle} (but with different conformal dimensions), and both gravity and gauge fluctuations are introduced in the same way as before. In other words, the (0+1)-dimensional complex SYK model can be thought of as defined on the boundary of the near-horizon AdS$_2$ region of the AdS$_4$ black brane.

\begin{figure}[htb!]
\begin{center}
\includegraphics[width=11cm, angle=0]{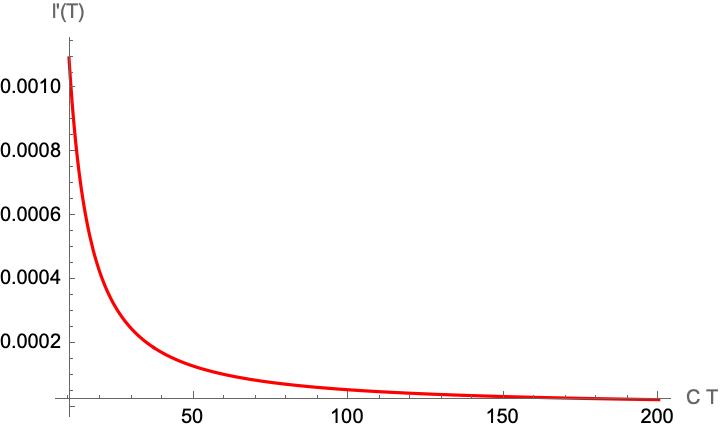}
\caption{The renormalized parameter $\ell'$ as a function of $C T$ (with a special choice of parameters $\ell=1/4$, $q=1$, $K=10^4$, and $\omega=10^{-3}$)}\label{fig:sykdcc1}
\end{center}
\end{figure}

\begin{figure}[htb!]
\begin{center}
\includegraphics[width=11cm, angle=0]{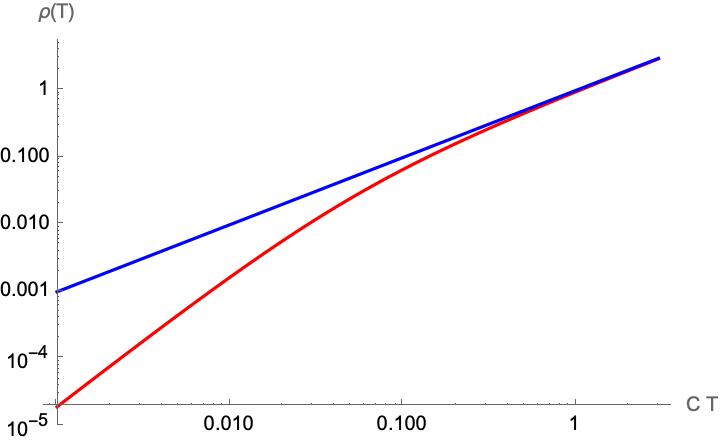}
\caption{The uncorrected (blue) and quantum-corrected (red) DC resistivities as functions of $C T$ (with a special choice of parameters $\ell=1/4$, $q=1$, $K=10^4$, and $\omega=10^{-3}$)}\label{fig:sykdcc2}
\end{center}
\end{figure}

Plugging the quantum-corrected Green's function back into Eq.~\eqref{eq:sykdc}, we obtain the DC resistivity with quantum corrections from the (0+1)-dimensional complex SYK model:
\begin{equation}
\rho'_{dc}\,\, \propto\,\, T^{1-2\ell'}\, ,
\end{equation}
where the quantum-corrected renormalized parameter $\ell'$ is shown in Fig.~\ref{fig:sykdcc1}, and the quantum-corrected DC resistivity is shown in Fig.~\ref{fig:sykdcc2}. We find when $T < C^{-1}$ the quantum fluctuations become dominant, and the quantum correction is qualitatively different from the one in the AdS$_4$/CFT$_3$ approach (see Fig.~\ref{fig:dcc1}) due to the different temperature dependences of DC resistivity in these two models. In actual experiments, both formalisms could play a role in certain circumstances, depending on the specific properties of materials.

\subsection{Optical conductivity}

For optical conductivity, we consider the regime $\omega\gg T$. The optical conductivity is given by \cite{Michon:2023qpl, PhysRevB.92.054305}:
\begin{equation}\label{eq:sykoptical}
\sigma(\omega)=\frac{iH}{\omega}\int d\epsilon\frac{f(\epsilon)-f(\epsilon+\omega)}{\omega+\Sigma^{*}(\epsilon)-\Sigma(\epsilon+\omega)}\, ,
\end{equation}
where $H$ is a positive constant, and $f(\epsilon)=(e^{\epsilon/T}+1)^{-1}$ is the Fermi-Dirac function. In the limit $\omega\gg T$, the self-energy has the scaling form
\be
\Sigma(\omega)=c\, \omega^{1-2\ell}\, ,
\ee
where $c$ is a constant independent of $\omega$ and $T$. Then, the optical conductivity becomes
\begin{equation}
\sigma(\omega)=\frac{iH}{\omega}\int d\epsilon\frac{f(\epsilon)-f(\epsilon+\omega)}{\omega+c^*\epsilon^{1-2\ell}-c(\epsilon+\omega)^{1-2\ell}}\, .
\end{equation}
In terms of new variables $b\equiv \epsilon / T$ and $s\equiv \omega / T$, we can express $\sigma (\omega)$ as
\begin{equation}
\sigma(\omega)=\frac{iH}{s}\int db\frac{(e^b+1)^{-1}-(e^{b+s}+1)^{-1}}{sT+c^*b^{1-2\ell}T^{1-2\ell}-c(b+s)^{1-2\ell}T^{1-2\ell}}\, .
\end{equation}
Furthermore, we introduce a variable $y$ defined by $b = s (y-1) / 2$. Then,
\begin{equation}\label{eq:sigma intermediate}
\sigma(\omega)=\frac{iH}{s}\int^{1}_{-1}\frac{s}{2}dy\frac{(e^{\frac{s}{2}(y-1)}+1)^{-1}-(e^{\frac{s}{2}(y+1)}+1)^{-1}}{sT+c^{*}(\frac{s}{2})^{1-2\ell}(y-1)^{1-2\ell}T^{1-2\ell}-c(\frac{s}{2})^{1-2\ell}(y+1)^{1-2\ell}T^{1-2\ell}}\, .
\end{equation}
Since $\omega\gg T$, we have $s\equiv \omega / T\to\infty$, under which the integral \eqref{eq:sigma intermediate} becomes
\begin{align}
\sigma(\omega)&=\frac{iH}{2}\int^{1}_{-1}dy\frac{1}{sT+c^{*}(\frac{s}{2})^{1-2\ell}(y-1)^{1-2\ell}T^{1-2\ell}-c(\frac{s}{2})^{1-2\ell}(y+1)^{1-2\ell}T^{1-2\ell}}\nonumber\\
&=(sT)^{2\ell-1}\frac{iH}{2}\int^{1}_{-1}dy\frac{1}{(sT)^{2\ell}+c^{*}(\frac{1}{2})^{1-2\ell}(y-1)^{1-2\ell}-c(\frac{1}{2})^{1-2\ell}(y+1)^{1-2\ell}}\nonumber\\
&=\omega^{2\ell-1}\frac{iH}{2}\int^{1}_{-1}dy\frac{1}{\omega^{2\ell}+c^{*}(\frac{1}{2})^{1-2\ell}(y-1)^{1-2\ell}-c(\frac{1}{2})^{1-2\ell}(y+1)^{1-2\ell}}\, .
\end{align}
To obtain the scaling behavior of optical conductivity, we define a new function:
\begin{equation}
F(\ell,\omega) \equiv \int^{1}_{-1}dy\frac{1}{\omega^{2\ell}+c^{*}(\frac{1}{2})^{1-2\ell}(y-1)^{1-2\ell}-c(\frac{1}{2})^{1-2\ell}(y+1)^{1-2\ell}}\, .
\end{equation}
Consequently, the optical conductivity can be rewritten as
\begin{equation}
  \sigma(\omega)=\frac{iH}{2}\omega^{2\ell-1+\frac{{\rm log}F(\ell,\omega)}{{\rm log}\omega}}\, .
\end{equation}
We recall that in order to recover linear resistivity, we should take $\ell=0$ at the tree level. Applying it to the optical conductivity, we obtain the leading-order scaling behavior of optical conductivity $\sigma(\omega)\sim\omega^{-1}$.

\begin{figure}[htb!]
\begin{center}
\includegraphics[width=11cm, angle=0]{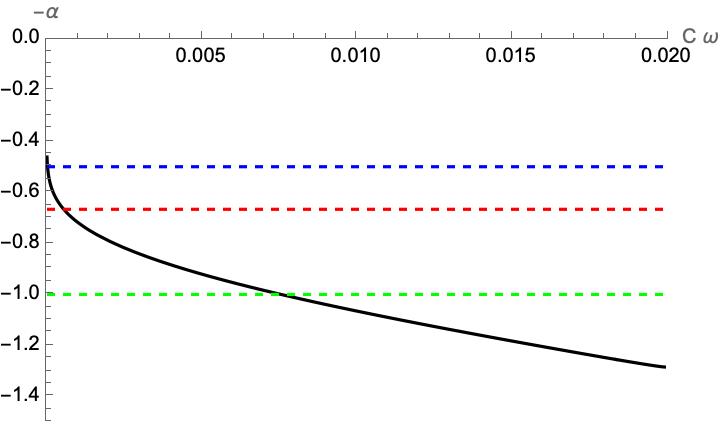}
\caption{The quantum-corrected $-\alpha$ as a function of $C\omega$ (with a special choice of parameters $\ell=0$, $q=1$, $K=1$, and $\mu=0$). The blue, red, and green dashed lines denote the special values $-1/2$, $-2/3$, and $-1$, respectively.}\label{fig:sykoptical}
\end{center}
\end{figure}

Because the quantum-corrected complex fermion Green's function formally remains the same as \eqref{eq:Green Fct with F3} in the AdS$_4$/CFT$_3$ approach (with different conformal dimensions), to introduce quantum correction, we consider the expansion of the quantum-corrected Green's function \eqref{eq:Green Fct with F3} at low temperature for small $\ell$ and small $\omega$. At the leading order in this limit,
\begin{equation}
\langle\mathcal{G}(\omega)\rangle=\frac{\ell}{(2C)^{2\ell}}\frac{{\rm coth}(\sqrt{2}\pi\sqrt{C(\omega-\frac{1}{2K}+\mu)})}{C(\omega-\frac{1}{2K}+\mu)}\, .
\end{equation}
Similar to the previous treatment, we define a new parameter $\ell'$ such that
\be
  \langle\mathcal{G}(\omega)\rangle = \frac{\ell}{(2C)^{2\ell}}\, \omega^{2\ell'-1}\, .
\ee
Then, $\ell'$ can be expressed as
\begin{equation}
\ell'=\frac{1}{2}+\frac{{\rm log} \left(\frac{{\rm coth}(\sqrt{2}\pi\sqrt{C(\omega-\frac{1}{2K}+\mu)})}{C(\omega-\frac{1}{2K}+\mu)}\right)}{2{\rm log}\omega}\, .
\end{equation}
Plugging it back into Eq.~\eqref{eq:sykoptical} and repeating the same steps, we obtain the quantum-corrected optical conductivity from the (0+1)-dimensional complex SYK model:
\begin{equation}
  \sigma'(\omega)=\frac{iH}{2}\omega^{2\ell'-1+\frac{{\rm log}F(\ell',\omega)}{{\rm log}\omega}}\, .
\end{equation}
The scaling power $-\alpha$ of $\sigma'(\omega)\, \propto\, \omega^{-\alpha}$ is shown in Fig.~\ref{fig:sykoptical}. We see that one can recover the experimental results for optical conductivity mentioned in the introduction at some special scales and parameters.

\section{Discussion}\label{sec:Discussion}

In this manuscript, we have explored quantum corrections' role in the holographic description of strange metals. The holographic description is based on a near-extremal AdS$_4$ black hole whose near-horizon region develops an infinite AdS$_2$ throat. The pattern of symmetry breaking for very low temperatures studied in the contexts of asymptotically AdS$_2$ JT gravity is quite relevant to the holographic treatment of the transport properties of strange metals, and we have considered how it modifies certain transport properties in the holographic approach.

Due to a factorization property of the near-horizon geometry, the DC and the optical conductivities of holographic strange metals are controlled by the IR fixed point, or more precisely, by the IR CFT$_1$ fermionic operator Green's function. It was, therefore,  sufficient for us to only consider the quantum corrections from gravity and gauge fluctuations in the IR CFT$_1$ fermionic operator Green's function. Within the framework of low-temperature corrections, we reproduced the previous holographic results for strange metal transport 
\cite{Faulkner:2013bna, Faulkner:2010zz, Faulkner:2009wj, Faulkner:2010da, Faulkner:2011tm, Liu:2009dm} in the limit where quantum fluctuations in the throat are neglected which also corresponds to considering the classical geometry of AdS$_2$. We went on to include perturbative temperature corrections to those previous results, which led to modifications of the DC and optical conductivities. One of our main results is that although the quantum correction to the DC resistivity is too small for the current experimental accuracy, the theoretical results could be tested in future more precise experiments.

One significant result from our explorations is that for the optical conductivity at low frequencies, $\omega \gg T$, the theoretical results with quantum correction can cover all the observed values of scaling powers in experiments. In the available experimental data, the parameter $\alpha$ in $\sigma \sim \omega^{-\alpha}$ is typically $2/3$ as summarized in \cite{PhysRevB.58.11631, Marel2006ScalingPO, Phillips:2022nxs}, but some other values  have been reported experimentally:  0.5 \cite{PhysRevB.66.041104}, 0. 6 \cite{vandeMarel:2003wn}, 0.65 \cite{vandeMarel:2003wn}, 0.7 \cite{Baraduc1996InfraredCI}, 0.77 \cite{PhysRevB.49.9846}, and 0.8 \cite{Michon:2023qpl}.  As shown in Fig.~\ref{fig:optical}, we are able to fit all the experimental values of $\alpha$. 

Another prominent approach to the physics of strange metals has been rooted in models such as the (0+1)-dimensional complex SYK and the (1+1)-dimensional Yukawa-SYK quantum-mechanical models. In section 
\ref{sec:CplxSYK}, we compared our holographic AdS$_4$/CFT$_3$  approach with the complex SYK results. One immediate qualitative difference between the two approaches arises from the form that the retarded Green's function, $G_R^{1d}$, enters in determining transport. In the holographic approach, given that we consider fluctuations in the throat as embedded in the larger, higher-dimensional spacetime,  AdS$_2\in$ AdS$_4$, we have that $G_R^{3d} \sim  (G^{1d}_R)^{-1}$. In direct models like the complex SYK, transport properties are directly determined by $G_R^{1d}$. Consequently, the AdS$_4$/CFT$_3$ and the (0+1)-dimensional complex SYK model lead to qualitatively different behaviors of the DC conductivities.

\subsection{A sweet spot scenario for quantum corrections}

In the spirit of potential applications, let us estimate the values of temperature below which we expect quantum corrections to be substantial. Based on the analyses in \cite{Larsen:2019oll, Nian:2020qsk, David:2020jhp, Iliesiu:2020qvm}, the quantum temperature $T_q$ can be estimated using the heat capacity of the AdS$_4$ black hole (or black brane). Since the entropy of a near-extremal AdS$_4$ black hole has the following expression with a quantum correction term:
\begin{align}
  S & = S_0 + 4 \pi^2 C T + \frac{3}{2}\, \textrm{log} (C T) \nonumber\\
  {} & = S_0 + \left(\frac{C_p}{T} \right)_{T=0}\cdot T + \frac{3}{2}\, \textrm{log} (C T)\, ,
\end{align}
where $S_0$ is the AdS$_4$ black hole entropy in the extremal limit, and $C_p$ denotes the heat capacity. In the literature, a metal's heat capacity is linear in $T$ at low temperatures. The proportionality constant $\gamma \equiv C_p / T$, also called the Sommerfeld coefficient, has recently been measured for a class of strange metal (Ba$_4$Nb$_{1-x}$Ru$_{3+x}$O$_{12}$, $|x| < 0.20$), which takes values in the range $[164\, \textrm{mJ}/(\textrm{mol}\cdot \textrm{K}^2),\, 275\, \textrm{mJ}/(\textrm{mol}\cdot \textrm{K}^2)]$ for $T \in [50\, \textrm{mK},\, 30\, \textrm{K} ]$ \cite{PhysRevLett.132.226503}. Based on these data, an estimate of the quantum temperature in this case is $T_q \equiv 1/\gamma \simeq [30\, \textrm{K},\, 50\, \textrm{K}]$, which, in principle, allows an experimental detection of quantum corrections.

\begin{figure}[htb!]
	\centering
	\subfigure[correction for (2+1)-d model]{
		\begin{minipage}[b]{0.45\textwidth}
			\includegraphics[width=1\textwidth]{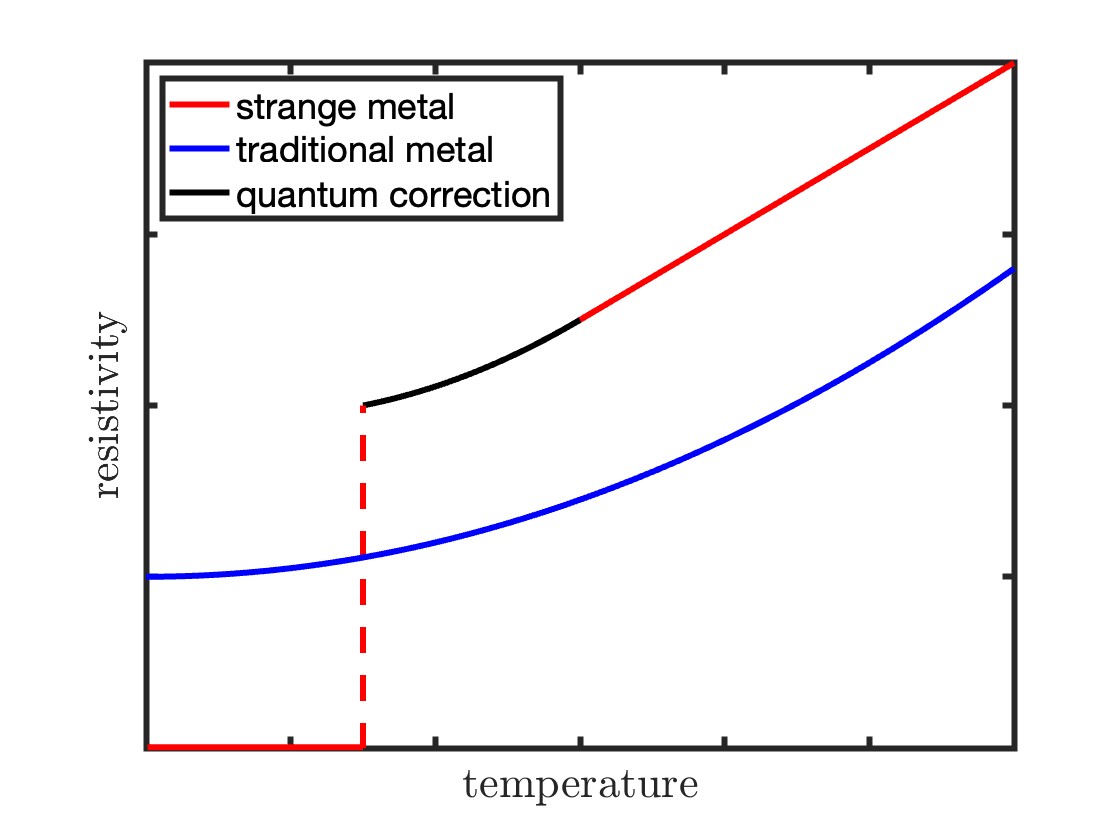} 
		\end{minipage}
	}
    	\subfigure[correction for (0+1)-d model]{
    		\begin{minipage}[b]{0.45\textwidth}
   		 	\includegraphics[width=1\textwidth]{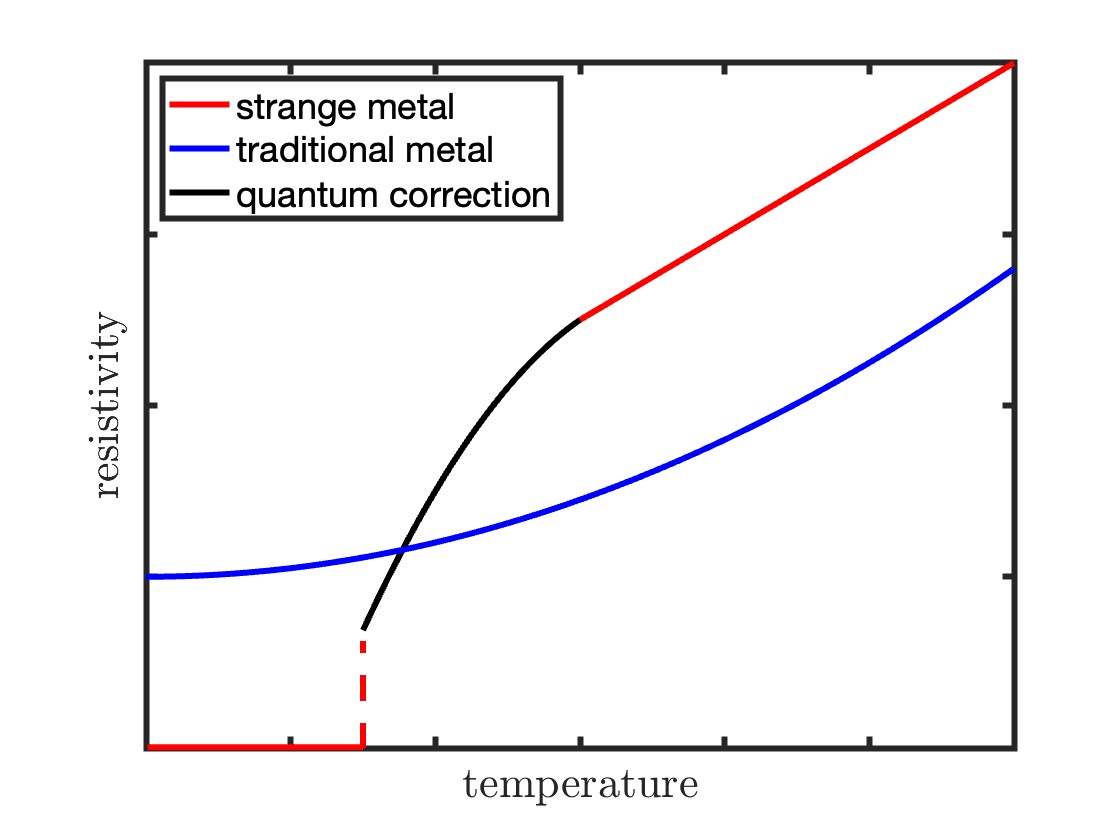}
    		\end{minipage}
    	}
	\\ 
	\subfigure[phase diagram]{
		\begin{minipage}[b]{0.55\textwidth}
			\includegraphics[width=1\textwidth]{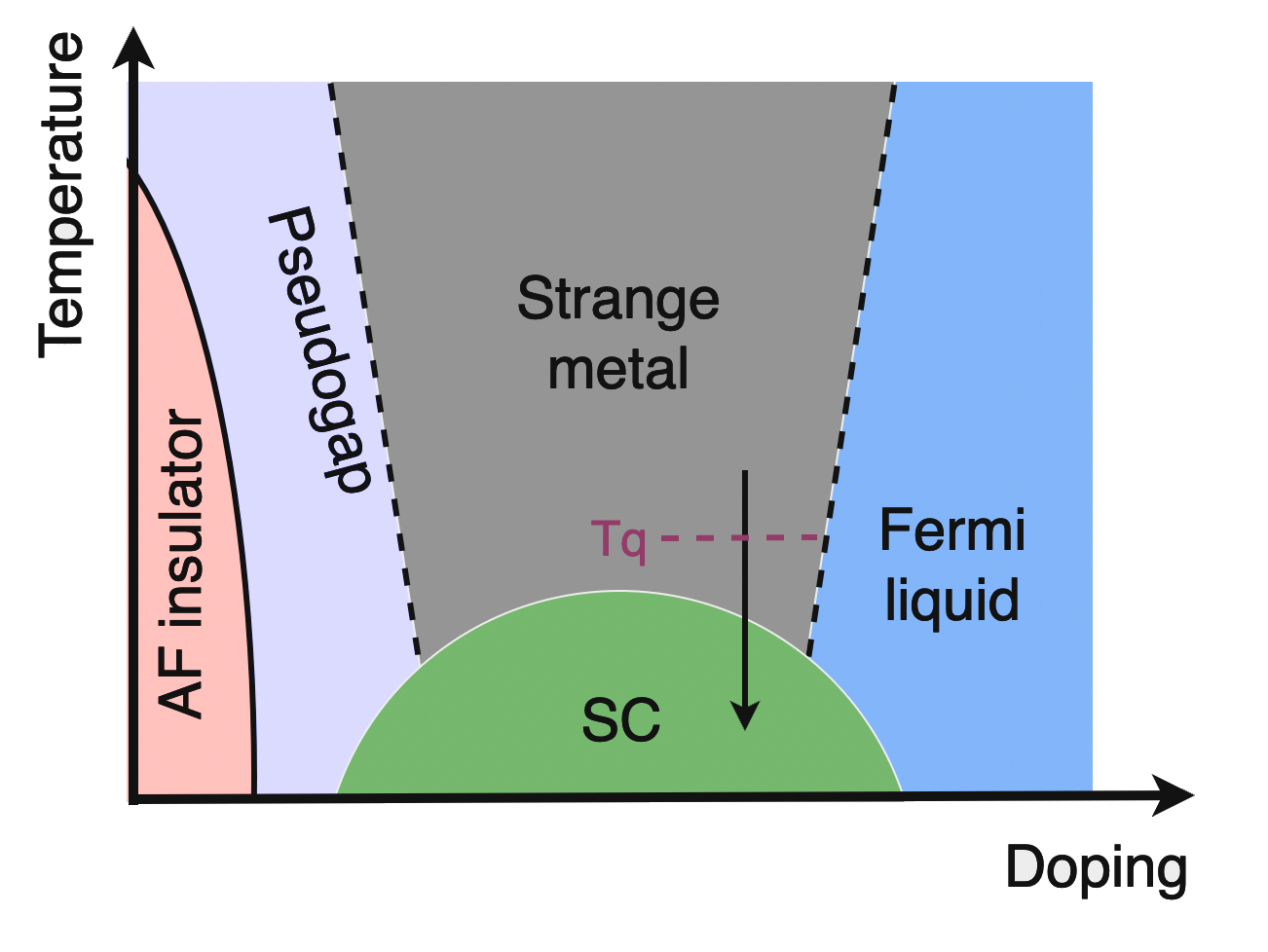} 
		\end{minipage}
	}
	\caption{The fourth possible scenario of strange metal's resistivity in materials}\label{fig:Scenario 4}
\end{figure}

According to our theoretical prediction, the deviation from the DC linear resistivity should manifest below the temperature $T_q \simeq \gamma^{-1}$. However, we should also take into account the transition from strange metal to other phases. Due to the competition between $T_q$ and the critical temperature $T_c$, different scenarios could be observed in various materials. We are going to concentrate on the sweet spot scenario, where quantum corrections are most pronounced, with details presented in Fig.~\ref{fig:Scenario 4}.

In this scenario, the material transfers from a strange metal phase to a superconducting phase when the temperature is lowered. However, the quantum temperature $T_q$ is higher than the transition temperature $T_c$, which makes the quantum corrections to the linear DC resistivity manifest while still in the strange metal phase (see Fig.~\ref{fig:Scenario 4}). Depending on the quasi-dimension of the material, we should apply either the AdS$_4$/CFT$_3$ or the SYK-like model description of the strange metal, which leads to different behaviors ((a) and (b) shown in Fig.~\ref{fig:Scenario 4}).

There are other scenarios where the effects of quantum corrections will not be observed, for example, if the scale of quantum corrections, $T_q$, is below the transition temperature to the superconducting phase, $T_c$, or in cases where the doping is such that the strange metal phase transitions to the Fermi liquid phase.

Let us conclude by listing a number of open questions that our work stimulates. It would be natural to consider the Kerr-Newman black brane to see how the rotation will affect the conductivities of strange metal. Finding a satisfactory theoretical explanation of strange metal's Hall angle \cite{PhysRevLett.67.2092, PhysRevLett.67.2088, Ahn2023InabilityOL, PhysRevLett.69.2855} is another open question. We may consider a dyonic black hole and see if quantum corrections clarify various issues surrounding the Hall angle.

\section*{Acknowledgments}

We would like to thank Finn Larsen,  Li Li, Xin Meng, Ren\'e Meyer, Maulik Sabyasachi, Francesco Sannino, Yawen Sun, Jinwu Ye, Congyuan Yue, Hongbao Zhang, and Jingchao Zhang for many helpful comments and discussions. This work is supported in part by the NSFC under grants No.~12375067, No.~12147103, and No.~12247103. LPZ is partially supported by the U.S. Department of Energy under grant DE-SC0007859.  J.N. would like to thank the International Centre for Theoretical Physics (ICTP) and the Niels Bohr Institute (NBI) for their warm hospitality during the final stage of this work.

\newpage
\appendix

\section{Quantum averaged Green's function from gravity and gauge fluctuations}\label{Sec:AppendixA}

To evaluate the quantum averaged Green's function $\langle G_f (\omega) \rangle$ from gravity fluctuations, \eqref{eq:g2}, we use a saddle-point approximation. In this appendix, we present some details of the computation.

First, we rewrite the $M$-dependent terms in Eq.~\eqref{eq:g2} as a Gaussian integral:
\be\label{eq:Gaussian}
  \int dM\, e^{2\pi\sqrt{2CM} - M\beta + {\rm ln}\, \Gamma\left(\ell + i\frac{\omega}{2}\sqrt{\frac{2C}{M}}\right)+{\rm ln}\, \Gamma\left(\ell-i\frac{\omega}{2}\sqrt{\frac{2C}{M}}\right)+(\ell-\frac{1}{2})\, {\rm ln}\, 8CM} = \int dM\, e^{f(M)}\, ,
\ee
where
\be
f(M) = 2\pi\sqrt{2CM}-M\beta+{\rm ln}\Gamma\left(\ell+i\frac{\omega}{2}\sqrt{\frac{2C}{M}}\right) + {\rm ln}\Gamma\left(\ell-i\frac{\omega}{2}\sqrt{\frac{2C}{M}}\right) + \left(\ell-\frac{1}{2}\right){\rm ln}(8CM)\, .\label{eq:a2}
\ee
The saddle points are the solutions to the following equation:
\begin{align}
0 = \frac{\partial f(M)}{\partial M} & = \pi\sqrt{\frac{2C}{M}}-\beta-i\frac{\sqrt{2C}\omega}{4}M^{-3/2}\psi \left(\ell + i\frac{\sqrt C\omega}{\sqrt{M}} \right)\nonumber\\
& \quad + i\frac{\sqrt{2C}\omega}{4}M^{-3/2}\psi\left(\ell-\textcolor{black}{i\frac{C\omega}{\sqrt{M}}}\right)+\frac{\ell-\frac{1}{2}}{M}\, ,
\end{align}
where $\psi(x)$ is the digamma function, $\psi(z)=\frac{d}{dz}\ln \Gamma(z)$. In the limit $\omega\ll M$, the leading-order saddle-point solutions are given by
\begin{equation}\label{eq:sqrtm}
\frac{1}{\sqrt{2CM}}=\frac{-\pi\pm\sqrt{\pi^2+2(2\ell-1)\frac{\beta}{2C}}}{2\ell-1}\, .
\end{equation}
We expand the solution for positive $\sqrt{C M}$ and small $\beta$:
\begin{align}\label{eq:solution1}
\frac{1}{\sqrt{2CM}} & =\frac{-\pi+\sqrt{\pi^2+2(2\ell-1)\frac{\beta}{2C}}}{2\ell-1} = \frac{1}{\pi}\frac{\beta}{2C} + \frac{(1-2\ell)}{2\pi^3} \left(\frac{\beta}{2C}\right)^2 + O(\beta^3)\, .
\end{align}
Plugging it back into Eq.~\eqref{eq:Gaussian}, we have
\begin{align}
{} & \langle \mathcal{G}_f(\omega)\rangle\nonumber\\
 =\, & \frac{\sqrt{\pi} \left(\frac{\beta}{2\pi^2C}\right)^{3/2}}{4e^{\frac{2\pi^2C}{\beta}}(2C)^{2\ell-2}}\frac{1}{\Gamma(2\ell)}\int dM e^{2\pi\sqrt{2CM}-M\beta}\, \left(2\sqrt{2CM}\right)^{2\ell-1}\cdot \Gamma \left(\ell+i\frac{\omega}{2}\sqrt{\frac{2C}{M}}\right)\, \Gamma \left(\ell-i\frac{\omega}{2}\sqrt{\frac{2C}{M}}\right)\nonumber\\
 =\, & \frac{\sqrt{\pi} \left(\frac{\beta}{2\pi^2C}\right)^{3/2}}{4e^{\frac{2\pi^2C}{\beta}}(2C)^{2\ell-2}}\frac{1}{\Gamma(2\ell)}\int dM e^{2\pi\sqrt{2CM}-M\beta}\, \left(2\sqrt{2CM}\right)^{2\ell-1}\cdot \frac{\pi}{{\rm sin} \left(\pi\ell+i\pi\frac{\omega}{2}\sqrt{\frac{2C}{M}}\right)}\frac{\Gamma \left(\ell-i\frac{\omega}{2}\sqrt{\frac{2C}{M}}\right)}{\Gamma \left(1-\ell-i\frac{\omega}{2}\sqrt{\frac{2C}{M}}\right)}\nonumber\\
 =\, & \frac{\sqrt{\pi} \left(\frac{\beta}{2\pi^2C}\right)^{3/2}}{4\textcolor{black}{e^{\frac{2\pi^2C}{\beta}}}(2C)^{2\ell-2}}\, \frac{(2\pi)^{2\ell-1}}{\Gamma(2\ell)}\, \textrm{exp}\left[\frac{1+\frac{1-2\ell}{\pi^2}\frac{\beta}{2C}}{\left(\frac{1}{\pi}+\frac{1-2\ell}{2\pi^3}\frac{\beta}{2C}\right)^2\frac{\beta}{2C}}\right]\cdot \left[\frac{\beta}{2C}+\frac{1-2\ell}{2\pi^2} \left(\frac{\beta}{2C}\right)^2\right]^{1-2\ell}\nonumber\\
 {} & \times \frac{\pi}{{\rm sin} \left(\pi\ell + iC\omega \left[\frac{\beta}{2C}+\frac{(1-2\ell)}{2\pi^2} \left(\frac{\beta}{2C}\right)^2\right]\right)}\cdot \frac{\Gamma \left(\ell-i\frac{C\omega}{\pi} \left[\frac{\beta}{2C} + \frac{(1-2\ell)}{2\pi^2} \left(\frac{\beta}{2C}\right)^2\right]\right)}{\Gamma(1-\ell-i\frac{C\omega}{\pi}[\frac{\beta}{2C}+\frac{(1-2\ell)}{2\pi^2}(\frac{\beta}{2C})^2])}\nonumber\\
 \approx\, & \frac{\sqrt{\pi} \left(\frac{\beta}{2\pi^2C}\right)^{3/2}}{4(2C)^{2\ell-2}} \frac{(2\pi)^{2\ell-1}}{\Gamma(2\ell)} \left[\frac{\beta}{2C}+\frac{1-2\ell}{2\pi^2} \left(\frac{\beta}{2C}\right)^2\right]^{1-2\ell}\cdot \frac{\pi}{{\rm sin} \left(\pi\ell + iC\omega \left[\frac{\beta}{2C}+\frac{(1-2\ell)}{2\pi^2} \left(\frac{\beta}{2C}\right)^2\right]\right)}\nonumber\\
 & \times \frac{\Gamma \left(\ell - i\frac{C\omega}{\pi} \left[\frac{\beta}{2C}+\frac{(1-2\ell)}{2\pi^2} \left(\frac{\beta}{2C}\right)^2\right]\right)}{\Gamma \left(1 - \ell - i\frac{C\omega}{\pi} \left[\frac{\beta}{2C}+\frac{(1-2\ell)}{2\pi^2} \left(\frac{\beta}{2C}\right)^2\right]\right)}\, ,\label{app eq:c1}
\end{align}
where in the second equality, we have used the identity
\be
  \Gamma\left(\ell+i\frac{\omega}{2}\sqrt{\frac{2C}{M}}\right)\, \Gamma \left(\ell-i\frac{\omega}{2}\sqrt{\frac{2C}{M}}\right) =  \frac{\pi}{{\rm sin} \left(\pi\ell + i\frac{\omega}{2}\sqrt{\frac{2C}{M}}\right)}\frac{\Gamma \left(\ell-i\frac{\omega}{2}\sqrt{\frac{2C}{M}}\right)}{\Gamma \left(1-\ell-i\frac{\omega}{2}\sqrt{\frac{2C}{M}}\right)}\, .
\ee

\newpage
\subsection{Quantum corrections from gauge fluctuations}

We start with $\langle \mathcal{G}_{\tilde{\Lambda}}\rangle$ defined in \eqref{eq:correction2}:

\begin{align}
\langle \mathcal{G}_{\tilde{\Lambda}}\rangle & = e^{-\frac{\tau(\beta-\tau)}{2K\beta}} e^{\mu\tau}\frac{\theta_3 \left(i\frac{2\pi K}{\beta},-\frac{2\pi K}{\beta}\frac{\mu\beta}{2\pi}-\frac{\tau}{\beta}\right)}{\theta_3 \left(i\frac{2\pi K}{\beta},-\frac{2\pi K}{\beta}\frac{\mu\beta}{2\pi}\right)}\nonumber\\
& = e^{-\frac{\tau(\beta-\tau)}{2K\beta}} e^{\mu\tau}\frac{\sum_{m} e^{-i(2\pi m)\frac{\tau}{\beta}}e^{-\pi\frac{2\pi K}{\beta}m^2}e^{-K\frac{(i\mu\beta 2\pi m)}{\beta}}}{\sum_{m} e^{-\pi\frac{2\pi K}{\beta}m^2} e^{-K\frac{(i\mu\beta 2\pi m)}{\beta}}}\nonumber\\
& \approx e^{-\frac{\tau(\beta-\tau)}{2K\beta}}e^{\mu\tau}
\left(1+\sum_{m=1}e^{-i(2\pi m)\frac{\tau}{\beta}} e^{-\pi\frac{2\pi K}{\beta}m^2}e^{-K\frac{(i\mu\beta 2\pi m)}{\beta}}\right)\cdot \left(1-\sum_{n=1} e^{i(\pi i)\frac{2\pi K}{\beta}n^2} e^{-K\frac{(i\mu\beta 2\pi n)}{\beta}}\right)\nonumber\\
& = e^{-\frac{\tau(\beta-\tau)}{2K\beta}}e^{\mu\tau} \Bigg[1 -\sum_{n=1}e^{-\pi\frac{2\pi K}{\beta}n^2} e^{-K\frac{(i\mu\beta 2\pi n)}{\beta}} + \sum_{m=1}e^{-i(2\pi m)\frac{\tau}{\beta}} e^{-\pi\frac{2\pi K}{\beta}m^2} e^{-K\frac{(i\mu\beta 2\pi m)}{\beta}}\nonumber\\
& \qquad\qquad\qquad -\sum_{m=1}e^{-i(2\pi m)\frac{\tau}{\beta}} e^{-\pi\frac{2\pi K}{\beta}m^2} e^{-K\frac{(i\mu\beta 2\pi m)}{\beta}}\sum_{n=1} e^{-\pi\frac{2\pi K}{\beta}n^2} e^{-K\frac{(i\mu\beta 2\pi n)}{\beta}}\Bigg]\, ,\label{eq:gaugecorrection1}
\end{align}
where in the second equality, we used an expression for the theta function, $\theta_3$, as a sum, and in the third equality, we adopted an approximation for $\beta \to 0$.

\section{Some expressions in the limit $CT \ll 1$ for DC conductivity}\label{app:the limit CT << 1}

In Appendices \ref{app:the limit CT << 1} and \ref{app:optical conductivity}, we present the results in some limits not discussed in the main text. For simplicity, we only consider the quantum correction from gravity fluctuations.

\subsection{(2+1)-dimensional holographic model}

For the $CT\ll1$ case, one shall reconsider Eq.~\eqref{eq:g2}. Since $\ell$ is of the order of the particle mass and $CM$ is extremely small, we shall take the limit $\ell\gg CM$, then Eq.~\eqref{eq:g2} becomes
\begin{align}
\langle \mathcal{G}_f(t)\rangle & \approx \frac{\left(\frac{\beta}{2\pi^2C}\right)^{3/2}}{\sqrt{\pi}e^{\frac{2\pi^2C}{\beta}}(2C)^{2\ell-2}}\frac{\Gamma^2(\ell)}{\Gamma(2\ell)}\int \frac{1}{16}dM e^{4\pi\sqrt{2CM}}e^{-M\beta}\nonumber\\
{} & \quad \times\Gamma \left(\ell+i\left(\frac{\omega}{2}\sqrt{\frac{2C}{M}}\right)\right)\, \Gamma \left(\ell-i\left(\frac{\omega}{2}\sqrt{\frac{2C}{M}}\right)\right)\, .\label{eq:b1}
\end{align}
We rewrite the $M$-dependent terms as a Gaussian integral:
\be
  \int dM\, e^{4\pi\sqrt{2CM} - M\beta + {\rm ln}\, \Gamma\left(\ell + i\frac{\omega}{2}\sqrt{\frac{2C}{M}}\right)+{\rm ln}\, \Gamma\left(\ell-i\frac{\omega}{2}\sqrt{\frac{2C}{M}}\right)} = \int dM\, e^{f(M)}\, ,
\ee
where
\be
f(M) = 4\pi\sqrt{2CM}-M\beta+{\rm ln}\Gamma\left(\ell+i\frac{\omega}{2}\sqrt{\frac{2C}{M}}\right) + {\rm ln}\Gamma\left(\ell-i\frac{\omega}{2}\sqrt{\frac{2C}{M}}\right) \, .
\ee
The saddle points are the solutions to the following equation:
\begin{align}
0 = \frac{\partial f(M)}{\partial M} & = 2\pi\sqrt{\frac{2C}{M}}-\beta-i\frac{\sqrt{2C}\omega}{4}M^{-3/2}\psi \left(\ell + i\frac{\sqrt C\omega}{\sqrt{M}} \right)\nonumber\\
&\quad + i\frac{\sqrt{2C}\omega}{4}M^{-3/2}\psi\left(\ell-i\frac{C\omega}{\sqrt{M}}\right)\, .
\end{align}
For $\omega\ll M$, the leading-order saddle-point solutions are given by
\begin{equation}
0 = 2\pi\sqrt{\frac{2C}{M}}-\beta \Rightarrow \frac{1}{\sqrt{2CM}}=\frac{1}{2\pi}\frac{\beta}{2C} \, .
\end{equation}
Plugging it back to Eq.~\eqref{eq:b1}, we obtain
\begin{align}
{} & \langle \mathcal{G}_f(\omega)\rangle\nonumber\\
 =&\frac{\left(\frac{\beta}{2\pi^2C}\right)^{3/2}}{16\sqrt{\pi}e^{\frac{2\pi^2C}{\beta}}(2C)^{2\ell-2}}\frac{\Gamma^2(\ell)}{\Gamma(2\ell)}\int dM e^{4\pi\sqrt{2CM}}e^{-M\beta}\cdot \Gamma \left(\ell+i\frac{\omega}{2}\sqrt{\frac{2C}{M}}\right)\, \Gamma \left(\ell-i\frac{\omega}{2}\sqrt{\frac{2C}{M}}\right)\nonumber\\
 =\, & \frac{\left(\frac{\beta}{2\pi^2C}\right)^{3/2}}{16\sqrt{\pi}e^{\frac{2\pi^2C}{\beta}}(2C)^{2\ell-2}}\frac{\Gamma^2(\ell)}{\Gamma(2\ell)}\int dM e^{4\pi\sqrt{2CM}}e^{-M\beta}\cdot \frac{\pi}{{\rm sin} \left(\pi\ell+i\pi\frac{\omega}{2}\sqrt{\frac{2C}{M}}\right)}\frac{\Gamma \left(\ell-i\frac{\omega}{2}\sqrt{\frac{2C}{M}}\right)}{\Gamma \left(1-\ell-i\frac{\omega}{2}\sqrt{\frac{2C}{M}}\right)}\nonumber\\
 \approx\, & \frac{\left(\frac{\beta}{2\pi^2C}\right)^{3/2}}{16\sqrt{\pi}e^{\frac{2\pi^2C}{\beta}}(2C)^{2\ell-2}}\frac{\Gamma^2(\ell)}{\Gamma(2\ell)} e^{\frac{8\pi^2C}{\beta}}  \cdot \frac{\pi}{{\rm sin} \left(\pi\ell+i\pi\frac{1}{4\pi}\frac{\omega}{T}\right)}\frac{\Gamma \left(\ell-i\frac{1}{4\pi}\frac{\omega}{T}\right)}{\Gamma \left(1-\ell-i\frac{1}{4\pi}\frac{\omega}{T}\right)}\nonumber\\
  =\, & \frac{\left(\frac{\beta}{2\pi^2C}\right)^{3/2}}{16\sqrt{\pi}(2C)^{2\ell-2}}\frac{\Gamma^2(\ell)}{\Gamma(2\ell)} e^{\frac{6\pi^2C}{\beta}}  \cdot \frac{\pi}{{\rm sin} \left(\pi\ell+i\pi\frac{1}{4\pi}\frac{\omega}{T}\right)}\frac{\Gamma \left(\ell-i\frac{1}{4\pi}\frac{\omega}{T}\right)}{\Gamma \left(1-\ell-i\frac{1}{4\pi}\frac{\omega}{T}\right)}\, . \label{eq:b6}
\end{align}
Thus, for the limit $\omega\ll T\ll\frac{1}{C}$, the temperature dependence of quantum-corrected Green's function \eqref{eq:b6} is 
\be
\langle \mathcal{G}_f(\omega)\rangle\sim T^{-3/2}.
\ee
The temperature dependence of quantum-corrected DC resistivity (without considering gauge fluctuations) in (2+1)-dimensional holographic model can be approximately written as 
\be
\rho'\sim T^{-3/2}.
\ee
Therefore, in this regime of parameters, the quantum fluctuations completely destroy the possibility of linear temperature resistivity.

\subsection{(0+1)-dimensional complex SYK model}

In this case, let us recall that to recover the linear resistivity at the classical level in the complex SYK model, we should take $\ell=0$, for which $\ell\ll CM$. Thus, the $f(M)$ equation is just the same as Eq.~\eqref{eq:a2} :
\be
f(M) = 2\pi\sqrt{2CM}-M\beta+{\rm ln}\Gamma\left(\ell+i\frac{\omega}{2}\sqrt{\frac{2C}{M}}\right) + {\rm ln}\Gamma\left(\ell-i\frac{\omega}{2}\sqrt{\frac{2C}{M}}\right) + \left(\ell-\frac{1}{2}\right){\rm ln}(8CM)\, .
\ee
For $\ell=0$,
\begin{align}
f(M) &= 2\pi\sqrt{2CM}-M\beta+{\rm ln}\Gamma\left(i\frac{\omega}{2}\sqrt{\frac{2C}{M}}\right) + {\rm ln}\Gamma\left(-i\frac{\omega}{2}\sqrt{\frac{2C}{M}}\right) - \frac{1}{2}{\rm ln}(8CM)\,\nonumber\\
&= 2\pi\sqrt{2CM}-M\beta+{\rm ln}\left( \frac{\pi}{\frac{\omega}{2}\sqrt{\frac{2C}{M}}{\rm sinh}(\pi\frac{\omega}{2}\sqrt{\frac{2C}{M}})}\right) - \frac{1}{2}{\rm ln}(8CM)\nonumber\\
&\simeq 2\pi\sqrt{2CM}-M\beta+{\rm ln}\left( \frac{2M}{\omega^2C}\right) - \frac{1}{2}{\rm ln}(8CM).
\end{align}
Then, the saddle-point equation is 
\begin{align}
0 = \frac{\partial f(M)}{\partial M}  &= \pi\sqrt{\frac{2C}{M}}-\beta+\frac{1}{M}-\frac{1}{2}\frac{1}{M}\,\nonumber \\
&=\pi\sqrt{\frac{2C}{M}} - \beta + \frac{1}{2}\frac{1}{M}.
\end{align}
The saddle-point solutions are given by
\be
\frac{1}{\sqrt{2CM}}=-\pi\pm\sqrt{\pi^2+2\frac{\beta}{2C}}\, .
\ee
We expand the solution for positive $\sqrt{C M}$ and large $\beta$:
\be
\frac{1}{\sqrt{2CM}}\simeq \sqrt{2}\sqrt{\frac{\beta}{2C}}.
\ee
Plugging it back into Eq.~\eqref{eq:g2}, we obtain
\begin{align}
  {} & \langle \mathcal{G}_f(\omega)\rangle \nonumber\\
\approx\,\, & \frac{\left(\frac{\beta}{2\pi^2C}\right)^{3/2}}{\sqrt{\pi}e^{\frac{2\pi^2C}{\beta}}(2C)^{2\ell-2}}\int \frac{1}{16}dM e^{4\pi\sqrt{2CM}}e^{-M\beta}\times \frac{1}{\Gamma(2\ell)}\, \Gamma \left(\ell+i(2\sqrt{2CM})\right)\, \Gamma \left(\ell-i(2\sqrt{2CM})\right)\nonumber\\
 {} & \times \Gamma \left(\ell+i\left(\frac{\omega}{2}\sqrt{\frac{2C}{M}}\right)\right)\, \Gamma \left(\ell-i\left(\frac{\omega}{2}\sqrt{\frac{2C}{M}}\right)\right)\nonumber\\
 =\,\, & \frac{\left(\frac{\beta}{2\pi^2C}\right)^{3/2}}{\sqrt{\pi}e^{\frac{2\pi^2C}{\beta}}(2C)^{2\ell-2}}\frac{1}{\Gamma(2\ell)}\int \frac{1}{16}dM e^{4\pi\sqrt{2CM}}e^{-M\beta}\cdot \frac{\pi}{2\sqrt{2CM}{\rm sinh}(2\sqrt{2CM})}\nonumber\\
 {} & \times\frac{\pi}{\frac{\omega}{2}\sqrt{\frac{2C}{M}}{\rm sinh} \left(\frac{\omega}{2}\sqrt{\frac{2C}{M}}\right)}\nonumber\\
  \approx\,\, & \frac{\left(\frac{\beta}{2\pi^2C}\right)^{3/2}}{\sqrt{\pi}e^{\frac{2\pi^2C}{\beta}}(2C)^{2\ell-2}}\frac{1}{\Gamma(2\ell)}\frac{1}{16}e^{4\pi\sqrt{\frac{C}{\beta}}-\frac{1}{2}}\cdot \frac{\pi}{2\sqrt{\frac{C}{\beta}}{\rm sinh}\left(2\sqrt{\frac{C}{\beta}}\right)}\cdot \frac{\pi}{\frac{\omega}{2} \sqrt{4C\beta}\, {\rm sinh} \left(\frac{\omega}{2}\sqrt{4C\beta} \right)} \nonumber\\
  \approx\,\, & \frac{\left(\frac{\beta}{2\pi^2C}\right)^{3/2} e^{-\frac{1}{2}}}{16\sqrt{\pi}(2C)^{2\ell}} \frac{1}{\Gamma(2\ell)} \frac{\pi^2}{4\omega^2}\, .\label{eq:b15}
\end{align}
Thus, the temperature dependence of the quantum-corrected Green's function \eqref{eq:b15} is 
\be
\langle \mathcal{G}_f(\omega)\rangle\sim T^{-3/2}.
\ee
Consequently, the temperature dependence of the quantum-corrected DC resistivity (without considering gauge fluctuations) in the (0+1)-dimensional complex SYK model is approximately
\be
\rho'\sim T^{3/2}.
\ee

\section{Some expressions in the limits $C \omega \gg 1$ and $C \omega \ll 1$ for optical conductivity}\label{app:optical conductivity}

\subsection{(2+1)-dimensional holographic model}

We start from low-temperature limit quantum-corrected Green's function \eqref{eq:c21} :
\be
\langle\mathcal{G}_f(\omega)\rangle= \frac{1}{4\pi^2(2C)^{2\ell}}\, \frac{1}{\Gamma(2\ell)}\, \left(e^{2\pi\sqrt{2C\omega}} - e^{-2\pi\sqrt{2C\omega}}\right)\cdot \Gamma^2(\ell+i\sqrt{2C\omega})\, \Gamma^2(\ell-2\sqrt{2C\omega})\, .\label{eq:appendixc1}
\ee

\begin{itemize}
\item In the limit $C\omega\gg1$, there is the identity 
\begin{equation}
\Gamma(\ell+i\sqrt{2C\omega})\Gamma(\ell-i\sqrt{2C\omega})=\frac{\pi \sqrt{2C\omega}}{{\rm sinh}\pi \sqrt{2C\omega}}\mathop{\overset{\ell-1}{\Pi}}_{n=1}(n^2+(\sqrt{2C\omega})^2).
\end{equation}
Let us recall in the (2+1)-dimensional holographic model, one shall take $\ell=1$ to recover the linear resistivity at the classical level. Thus, we have 
\begin{align}
\Gamma(\ell+i\sqrt{2C\omega})\Gamma(\ell-i\sqrt{2C\omega})&\approx\frac{\pi \sqrt{2C\omega}}{{\rm sinh}\pi \sqrt{2C\omega}}(\sqrt{2C\omega})^{2\ell-2}=\frac{\pi(\sqrt{2C\omega})^{2\ell-1}}{{\rm sinh}\pi \sqrt{2C\omega}}\nonumber\\
&\approx2\frac{\pi(\sqrt{2C\omega})^{2\ell-1}}{e^{\pi \sqrt{2C\omega}}}\, .
\end{align}
Plugging this back into  Eq.~\eqref{eq:appendixc1}, one has 
\be
\langle\mathcal{G}_f(\omega)\rangle= \frac{1}{(2C)^{2\ell}}\, \frac{1}{\Gamma(2\ell)}\,  (2C\omega)^{2\ell-1}\, .
\ee
This is just the tree-level zero-temperature limit of the Green's function up to some constants. The corresponding optical conductivity is given by 
\begin{equation}
\sigma(\omega)\;\propto\;\omega^{1-2\ell}.
\end{equation}
\item In the limit $C\omega\ll1$ which is compatible with $\sqrt{2C\omega}\ll\ell$, Eq.~\eqref{eq:appendixc1} becomes
\begin{align}
\langle\mathcal{G}_f(\omega)\rangle&= \frac{1}{4\pi^2(2C)^{2\ell}}\, \frac{1}{\Gamma(2\ell)}\, \left(e^{2\pi\sqrt{2C\omega}} - e^{-2\pi\sqrt{2C\omega}}\right)\cdot \Gamma^2(\ell+i\sqrt{2C\omega})\, \Gamma^2(\ell-2\sqrt{2C\omega})\,\nonumber\\
&\approx \frac{1}{4\pi^2(2C)^{2\ell}}\, \frac{1}{\Gamma(2\ell)}\, \left(4\pi\sqrt{2C\omega}\right)\cdot \Gamma^4(\ell)\, .
\end{align}
The corresponding quantum-corrected optical conductivity (without considering gauge fluctuations) is approximately 
\be
\sigma(\omega)\;\propto\;\omega^{-1/2},
\ee
which is consistent with the result shown in Fig.~\ref{fig:optical}.
\end{itemize}

\subsection{(0+1)-dimensional complex SYK model}

In the (0+1)-dimensional complex SYK model, one should take $\ell=0$ to recover the linear resistivity, thus $\ell\ll\sqrt{2C\omega}$. Then, Eq.~\eqref{eq:appendixc1} becomes
\begin{align}
\langle\mathcal{G}_f(\omega)\rangle&= \frac{1}{4\pi^2(2C)^{2\ell}}\, \frac{1}{\Gamma(2\ell)}\, \left(e^{2\pi\sqrt{2C\omega}} - e^{-2\pi\sqrt{2C\omega}}\right)\cdot \Gamma^2(\ell+i\sqrt{2C\omega})\, \Gamma^2(\ell-2\sqrt{2C\omega})\nonumber\\ 
&\approx\frac{1}{4\pi^2(2C)^{2\ell}}\, \frac{1}{\Gamma(2\ell)}\, \left(e^{2\pi\sqrt{2C\omega}} - e^{-2\pi\sqrt{2C\omega}}\right)\cdot 
4\frac{\pi^2(2C\omega)^{-1}}{e^{2\pi \sqrt{2C\omega}}}\, .\label{eq:appendixc8}
\end{align}

\begin{itemize}
\item In the limit $C\omega\gg1$. Eq.~\eqref{eq:appendixc8} becomes
\begin{align}
\langle\mathcal{G}_f(\omega)\rangle&=\frac{1}{4\pi^2(2C)^{2\ell}}\, \frac{1}{\Gamma(2\ell)}\, \left(e^{2\pi\sqrt{2C\omega}} - e^{-2\pi\sqrt{2C\omega}}\right)\cdot 
4\frac{\pi^2(2C\omega)^{-1}}{e^{2\pi \sqrt{2C\omega}}}\nonumber\\
&\approx\frac{1}{(2C)^{2\ell}}\, \frac{1}{\Gamma(2\ell)}\, (2C\omega)^{-1}\, .
\end{align}
The corresponding optical conductivity is given by
\be
\sigma(\omega)\;\propto\;\omega^{-1}\, .
\ee

\item In the limit $C\omega\ll1$. Eq.~\eqref{eq:appendixc8} becomes
\begin{align}
\langle\mathcal{G}_f(\omega)\rangle&=\frac{1}{4\pi^2(2C)^{2\ell}}\, \frac{1}{\Gamma(2\ell)}\, \left(e^{2\pi\sqrt{2C\omega}} - e^{-2\pi\sqrt{2C\omega}}\right)\cdot 
4\frac{\pi^2(2C\omega)^{-1}}{e^{2\pi \sqrt{2C\omega}}}\nonumber\\
&\approx \frac{1}{(2C)^{2\ell}}\, \frac{1}{\Gamma(2\ell)}\, 4\pi\left(2C\omega\right)^{-1/2}\, .
\end{align}
The corresponding quantum-corrected optical conductivity is approximately
\be
\sigma(\omega)\;\propto\;\omega^{-1/2}.
\ee
\end{itemize}

\bibliographystyle{utphys}
\bibliography{StrangeMetal}

\end{document}